\documentclass[journal]{IEEEtran}
\usepackage{amsmath}
\usepackage{amssymb}
\usepackage{amsfonts}
\usepackage{graphicx}
\usepackage{epsfig}
\usepackage{subfigure}
\usepackage{color}
\usepackage{amsthm}
\begin{document}

\title{Optimal Resource Allocation in Full-Duplex Wireless-Powered Communication Network}

\author{
        Hyungsik Ju and Rui Zhang,~\IEEEmembership{Member,~IEEE}        
        \thanks{This work was done when H. Ju was with the Department of Electrical and Computer Engineering, National University of Singapore. He is now with the Electronics and Telecommunications Research Institute, Korea (e-mail: jugun@etri.re.kr).}
        \thanks{R. Zhang is with the Department of Electrical and Computer Engineering, National University of Singapore (e-mail: elezhang@nus.edu.sg). He is also with the Institute for Infocomm Research, A*STAR, Singapore.}        
       }

\maketitle

\begin{abstract}
This paper studies optimal resource allocation in the wireless-powered communication network (WPCN), where one hybrid access-point (H-AP) operating in full-duplex (FD) broadcasts wireless energy to a set of distributed users in the downlink (DL) and at the same time receives independent information from the users via time-division-multiple-access (TDMA) in the uplink (UL). We design an efficient protocol to support simultaneous wireless energy transfer (WET) in the DL and wireless information transmission (WIT) in the UL for the proposed FD-WPCN. We jointly optimize the time allocations to the H-AP for DL WET and different users for UL WIT as well as the transmit power allocations over time at the H-AP to maximize the users' weighted sum-rate of UL information transmission with harvested energy. We consider both the cases with perfect and imperfect self-interference cancellation (SIC) at the H-AP, for which we obtain optimal and suboptimal time and power allocation solutions, respectively. Furthermore, we consider the half-duplex (HD) WPCN as a baseline scheme and derive its optimal resource allocation solution. Simulation results show that the FD-WPCN outperforms HD-WPCN when effective SIC can be implemented and more stringent peak power constraint is applied at the H-AP.
\end{abstract}

\begin{keywords}
Wireless-powered communication network (WPCN), wireless energy transfer (WET), full-duplex (FD) system, resource allocation, convex optimization.
\end{keywords}

\IEEEpeerreviewmaketitle

\setlength{\baselineskip}{1.0\baselineskip}
\newtheorem{definition}{\underline{Definition}}[section]
\newtheorem{fact}{Fact}
\newtheorem{assumption}{Assumption}
\newtheorem{theorem}{\underline{Theorem}}[section]
\newtheorem{lemma}{\underline{Lemma}}[section]
\newtheorem{corollary}{\underline{Corollary}}[section]
\newtheorem{proposition}{\underline{Proposition}}[section]
\newtheorem{example}{\underline{Example}}[section]
\newtheorem{remark}{\underline{Remark}}[section]
\newtheorem{algorithm}{\underline{Algorithm}}[section]
\newcommand{\mv}[1]{\mbox{\boldmath{$ #1 $}}}

\section{Introduction}
Traditionally, fixed energy sources (e.g. batteries) have been used to power energy-constrained wireless networks, such as sensor networks, which lead to limited operation time. Although the lifetime of wireless networks can be extended by replacing or recharging the batteries, it may be inconvenient, costly, and even dangerous (e.g., in a toxic environment) or infeasible (e.g., for sensors implanted in human bodies). As an alternative solution, energy harvesting (see e.g., \cite{Ozel}, \cite{Ho} and the references therein) has recently received a great deal of attention since it provides more cost-effective and truly perpetual energy supplies to wireless networks through scavenging energy from the environment.

Among other commonly used energy sources (e.g. solar and wind), radio signal radiated by ambient transmitters becomes a viable new source for energy harvesting. In particular, harvesting energy from the far-field radio-frequency (RF) signal transmission opens a new avenue for the unified study of wireless power transmission and wireless communication since radio signals carry energy and information at the same time. There are two main paradigms of research along this direction. One line of work focuses on studying the so-called simultaneous wireless information and power transfer (SWIPT) by characterizing the achievable trade-offs in simultaneous wireless energy transfer (WET) and wireless information transmission (WIT) with the same transmitted signal. SWIPT has recently been investigated for various channel setups, e.g., the point-to-point additive white Gaussian noise (AWGN) channel \cite{Grover}, \cite{Zhou}, the fading AWGN channel \cite{Liu}, \cite{Liu_DPS} the multi-antenna channel \cite{Zhang}-\cite{Ju_RBF}, the relay channel \cite{Fouladgar}, \cite{Nasir}, and the multi-carrier based broadcast channel \cite{Ng}-\cite{Zhou_OFDM}.

\begin{figure}[!t]
   \centering
   \includegraphics[width=0.8\columnwidth]{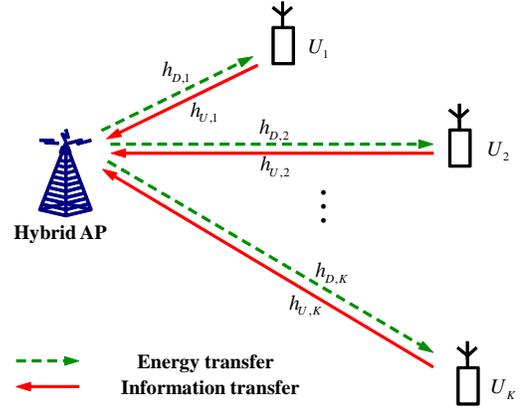}
   \caption{Wireless-powered communication network (WPCN) with DL WET and UL WIT.}
   \label{Fig_SystemModel}
\end{figure}

Another line of research is aimed to design a new type of wireless network termed wireless-powered communication network (WPCN) in which wireless terminals communicate using the energy harvested from wireless power transmissions. The WPCN has been studied under various network setups, such as cellular network \cite{Huang}, random-access network \cite{Nintanavongsa}, and multi-hop network \cite{Doost}. In \cite{Xie} and \cite{Ko}, another type of WPCN was investigated, where mobile charging vehicles are employed to wirelessly power the user devices in the network. In addition, the wireless-powered cognitive radio network was considered in \cite{Lee}, where active primary users are utilized as energy transmitters for charging their nearby secondary users. Furthermore, in our previous work \cite{Ju} we have studied one particular WPCN model, as shown in Fig. \ref{Fig_SystemModel}, where a hybrid access-point (H-AP) operating in time-division half-duplex (HD) mode coordinates WET/WIT to/from a set of distributed users in the downlink (DL) and uplink (UL) transmissions, respectively. It has been shown in \cite{Ju} that there exists a fundamental trade-off in allocating DL time for WET and UL time for WIT in the HD WPCN, since increasing DL time increases the amount of harvested energy and hence the UL transmit power at each user, but also decreases users' UL time for WIT given a total DL and UL time constraint.

On the other hand, there has been recently a growing interest in full duplex (FD) based wireless systems, where the wireless node transmits and receives simultaneously in the same frequency band, thus potentially doubling the spectral efficiency. However, due to the simultaneous transmission and reception at the same node, FD systems suffer from the self-interference (SI) that is part of the transmitted signal of a FD node received by itself, thus interfering with the desired signal received at the same time. Self-interference cancellation (SIC) is a key challenge for implementing FD communication since the power of SI typically overwhelms that of the desired signal. Various SIC techniques have been proposed in the literature (see e.g., \cite{Choi}-\cite{Bharadia} and the references therein), which are generally based on either analog-domain SIC (i.e., SIC in the wireless propagation channel or before the received signal is processed by analog-to-digital conversion (ADC)), digital-domain SIC (i.e., SIC after ADC by digital signal processing techniques), or their assorted combinations. By state-of-the-art SIC techniques today, it has been reported that SIC up to $110$dB higher power of the desired signal can be implemented \cite{Bharadia}. In addition, various effects of practical hardware limitations on FD communications, e.g., the finite dynamic range of transmit and/or receive filters \cite{Day}, \cite{Kim}, have been investigated. Furthermore, full-duplex techniques have been applied in various wireless communication applications for performance enhancement, e.g., spectral efficiency in relay network \cite{Ju-FDR}, multi-user cellular network \cite{Nguyen}, wireless physical layer security \cite{Zheng}, and cognitive radio network \cite{Heo}.

In this paper, we apply the FD technique to the WPCN shown in Fig. \ref{Fig_SystemModel}, to further improve throughput. We assume that the H-AP operates in the FD mode to broadcast energy and receive information to/from the distributed users simultaneously over a given frequency band, thus significantly saving the time for separate DL WET and UL WIT as compared to the HD-WPCN considered in \cite{Ju}. The proposed WPCN with the H-AP operating in FD mode is thus termed full-duplex WPCN (FD-WPCN), to differ from its HD-WPCN counterpart in \cite{Ju}. It is worth noting that the FD technique applied at the H-AP in our proposed FD-WPCN is for hybrid energy/communication transmission/reception, which is in sharp contrast to conventional setups with FD communications. However, similar to FD communication systems, the proposed hybrid energy/communication FD system is also subject to the practical issue of imperfect SIC at the H-AP, i.e., the interference due to the DL energy signal is not perfectly cancelled at the receiver for decoding the UL information.

The main results of this paper are summarized as follows:

\begin{itemize}
   \item For the proposed FD-WPCN, we present a new protocol to enable simultaneous WET in the DL and WIT in the UL over the same band. It is assumed that the H-AP operates in FD mode, while the users all operate in time division HD mode for the ease of implementation, which transmit independent information to the H-AP by time-division-multiple-access (TDMA) in the UL and harvest energy in the DL when they do not transmit. We also compare the proposed protocol for FD-WPCN with the harvest-then-transmit protocol proposed in \cite{Ju} for HD-WPCN.

   \item Under the proposed protocol for FD-WPCN, we characterize the maximum weighted sum-rate (WSR) of users in FD-WPCN, by jointly optimizing the time allocated to the H-AP for DL WET and users for UL WIT and the transmit power of the H-AP over time subject to a given total time constraint as well as the average and peak transmit power constraints at the H-AP. For the purpose of exposition, we first consider the ideal case of perfect SIC at the H-AP in FD-WPCN. It is shown that the WSR maximization problem in this case is convex, and hence we obtain the closed-form solution for the optimal time and power allocations by applying convex optimization techniques. It is revealed that the optimal time and power allocation is aligned with maximally exploiting the opportunistic communication gain in the network, i.e., more time is assigned to the users with stronger channels and/or higher rate weights (priorities) in UL WIT, while more power is broadcast by the H-AP in DL WET over time slots of the users with weaker channels and/or lower weights.

   \item We then study the practical case with imperfect SIC in FD-WPCN. In this case, the WSR maximization problem is shown to be non-convex, and thus is difficult to be solved optimally. Alternatively, we propose an efficient algorithm to find at least one locally optimal solution for this problem based on the optimal solution obtained for the ideal case without SI by iteratively optimizing the time and power allocations to maximize the WSR.

   \item Furthermore, we investigate the WSR maximization problem in a baseline HD-WPCN system based on the harvest-then-transmit protocol proposed in \cite{Ju}. It is worth pointing out the main difference between the optimization problems considered in this work and our prior work \cite{Ju} for HD-WPCN. In this paper, we study the joint power and time allocations subject to both the average and peak transmit power constraints in HD-WPCN, whereas only time allocation is considered in \cite{Ju} by assuming constant transmit power at the H-AP. By comparing the achievable rates of FD- versus HD-WPCNs, it is revealed that the former is more beneficial than the latter when the SI can be more effectively cancelled at the H-AP and/or the peak transmit power constraint at the H-AP is more stringent.
\end{itemize}

The rest of this paper is organized as follows. Section \ref{Sec:SystemModel} introduces the FD-WPCN model and the proposed transmission protocol. Section \ref{Sec:FD_WPCN} presents the optimal time and power allocation solutions to the WSR maximization problems in FD-WPCN for both the ideal case of perfect SIC and the practical case with imperfect SIC, respectively. Section \ref{Sec:HD_WPCN} addresses HD-WPCN, and shows the optimal solution for WSR maximization in this case. Section \ref{Sec:SimulationResults} presents simulation results for comparing the performance of FD-WPCN against HD-WPCN. Finally, Section VI concludes the paper.

\section{System Model}\label{Sec:SystemModel}
As shown in Fig. \ref{Fig_SystemModel}, this paper considers a WPCN with WET in the DL and WIT in the UL. The network consists of one H-AP and $K$ users (e.g., sensors) denoted by $U_i$, $i=1, \, \cdots, \, K$, operating over the same frequency band. It is assumed that users are sufficiently separated from each other. The H-AP is equipped with two antennas, and each user terminal is equipped with one antenna. The H-AP is assumed to have a stable energy supply, whereas each user terminal does not have any embedded energy sources. As a result, the users need to replenish energy from the received signal from the H-AP in the DL, which is then used to power operating circuit and transmit information in the UL. In this paper, we focus on the case of FD-WPCN where the H-AP operates in FD mode to broadcast energy for the DL WET and receive information for the UL WIT at the same time, while the users are assumed to all operate in time-division HD mode to harvest energy in the DL and transmit information in the UL orthogonally over time, for the ease of implementation. For the purpose of comparison, we also consider the case of HD-WPCN where the H-AP also operates in time-division HD mode, as studied in our previous work \cite{Ju}.

   \subsection{FD-WPCN}
   In FD-WPCN, the H-AP operates in FD mode by utilizing one antenna for transmitting energy to users in DL and the other antenna for receiving information from users in UL simultaneously over the same bandwidth. The DL channel from the H-AP to $U_i$ and the corresponding reversed UL channel are denoted by complex coefficients ${h_{D,i}}$ and ${h_{U,i}}$, $i=1, \,\, \cdots, \,\, K$, respectively. In addition, there exists a loopback channel at the H-AP, through which the transmitted DL energy signal of one antenna is received at the other antenna in addition to the UL information signal received from users. It is assumed that at the H-AP, analog domain SIC \cite{Choi}, \cite{Duarte2} is first performed,\footnote{We refer to SIC before A/D conversion as analog domain SIC, by e.g., offsetting the two transmit antennas by half a wavelength \cite{Choi}, or using an extra transmit RF chain to generate a reference RF signal \cite{Duarte2}, while SIC after A/D conversion, commonly referred to as digital domain SIC, will be considered later.} $g_A$ thus denotes the effective loopback channel after analog domain SIC at the H-AP. We also assume that $h_{D,i}$, $h_{U,i}$, $i = 1, \, \cdots, \, K$, and $g_A$ all follow quasi-static flat-fading, and the channels remain constant during each block transmission time, denoted by $T$, for the system of interest. It is further assumed that the H-AP knows perfectly $h_{U,i}$ and $h_{D,i}$, $i = 1, \,\, \cdots, \,\, K$.

   \begin{figure}[!t]
       \centering
       \includegraphics[width=0.8\columnwidth]{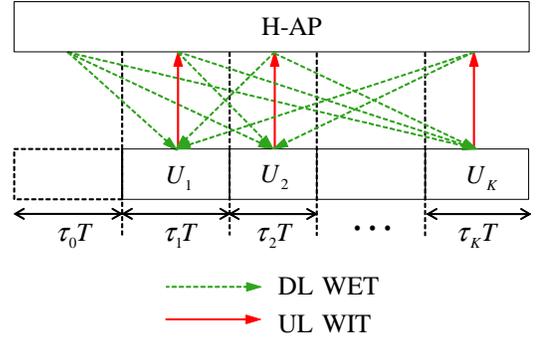}
       \caption{Simultaneous DL energy and UL information transmission in FD-WPCN.}
       \label{Fig_FD_WPCN_FrameStructure}
    \end{figure}

    In FD-WPCN, simultaneous WET in DL and WIT in UL can be achieved, as shown in Fig. \ref{Fig_FD_WPCN_FrameStructure}. Each transmit block is divided into $K+1$ slots each with duration of $\tau_i T$, $i = 0, \,\, 1, \,\, \cdots, \,\, K$. The $0\,$th slot is a dedicated power slot for DL WET only\footnote{This is to ensure that even for the case with one user in the system, i.e., $K=1$, the user can still receive energy from the dedicated power slot.}   and the $i\,$th slot, $i = 1, \,\, \cdots, \,\, K$, is used for both DL WET and UL WIT. During the $i\,$th slot, the H-AP broadcasts wireless energy with transmit power, denoted by $P_i$, to all users in the network. We thus have
   \begin{equation}\label{Eq_FD_SumTime}
      {\sum\limits_{i = 0}^{K} {{\tau _{i}}} \le 1.}
   \end{equation}
   In addition, the average transmit power of the H-AP over $K+1$ slots in each block and the peak transmit power of the H-AP at each slot are denoted by $P_{avg}$ and $P_{peak}$, respectively, i.e.,
   \begin{equation}\label{Eq_FD_AvgPower}
      {\sum\limits_{i = 0}^{K} {{\tau _{i}}P_i} \le P_{avg},}
   \end{equation}
   \begin{equation}\label{Eq_FD_PeakPower}
      {\,\,\,\,\,\,\,\,\,\,\,\,\,\,\,\,\,\,\,\,\,\,\,\,\,\,\,\,\,\,\,\,\,\,\,\,\,\,\,\,\,\,\,\,\,\,\,\, P_i \le P_{peak}, \,\, i = 0, \,\, \cdots, \,\, K.}
   \end{equation}
   In UL, we assume that the users transmit independent information to the H-AP using TDMA, i.e., the $i\,$th slot is allocated to user $U_i$ for WIT, $i = 1, \,\, \cdots, \,\, K$. Specifically, during the $i\,$th slot in each block, the H-AP broadcasts energy using one antenna and receives information from $U_i$ using the other antenna simultaneously, since the H-AP operates in FD mode. On the other hand, each user $U_i$, $i=1, \,\, \cdots, \,\, K$, transmits its own information to the H-AP in the $i\,$th slot, but cannot receive energy from the H-AP in the same slot, since it operates in time-division HD mode and has one single antenna. Instead, $U_i$ harvests energy from the H-AP during the other $K$ slots during which it is not scheduled for information transmission, and stores the energy harvested to be used in future.

   Consider the $j\,$th slot in one block of interest, $j = 0, \,\, \cdots, \,\, K$, during which the transmitted signal of the H-AP is denoted by $x_{A,j}$.\footnote{Note that $x_{A,j}$'s can also be used to send DL information (e.g., control signals from the H-AP to users) at the same time, which is not considered further in this paper. Interested readers may refer to the literature on SWIPT \cite{Liu}-\cite{Zhou_OFDM}.} We assume that $x_{A,j}$ is a pseudo-random sequence which is \emph{a prior} known at the H-AP\footnote{This facilities the implementation of SIC at the H-AP, as will be shown next.} satisfying $\mathbb E [ {{\left| {{x_{A,j}}} \right|}^2} ] = 1$. During the $j\,$th slot, the received signal at $U_i$, $i \ne j$, is then expressed as
   \begin{equation}\label{Eq_FD_ReceivedSignal_Ui}
      {{y_{i,j}} = \sqrt {{P_j}} {h_{D,i}}{x_{A,j}} + {z_{i,j}},}
   \end{equation}
   where ${y_{i,j}}$ and ${z_{i,j}}$ denote the received signal and noise at $U_i$, respectively. It is assumed that $P_j$ is sufficiently large such that the energy harvested due to the receiver noise is negligible. Furthermore, energy harvested due to the received UL WIT signals from other users is also assumed to be negligible since users are sufficiently separated from each other and their transmit power level is much lower as compared to $P_j$ in practice. Thus, the amount of energy harvested by $U_i$ during the $j\,$th slot can be expressed as
   \begin{equation}\label{Eq_FD_Harvested_Energy_Ui_Slot_j}
      {{E_{i,j}} = {\zeta _i}{P_j}{\left| {{h_{D,i}}} \right|^2}{\tau _j}T,}
   \end{equation}
   where $0 < \zeta_i < 1$, $i = 1, \,\, \cdots, \,\, K$, is the energy harvesting efficiency at $U_i$. For convenience, we assume $T = 1$ in the sequel of this paper without loss of generality. From (\ref{Eq_FD_Harvested_Energy_Ui_Slot_j}), the average harvested energy of $U_i$ in each block is thus expressed as
   \begin{equation}\label{Eq_FD_Harvested_Energy_Ui}
      {{E_{{U_i}}} = {\zeta _i}{\left| {{h_{D,i}}} \right|^2}\sum\limits_{\scriptstyle j = 0 \hfill \atop
      \scriptstyle j \ne i \hfill}^K {{\tau _j}{P_j}} .}
   \end{equation}

   On the other hand, during the $j\,$th slot with $j=i$, $U_i$ transmits its own information to the H-AP using a fixed portion of its average harvested energy per block given by (\ref{Eq_FD_Harvested_Energy_Ui}). Within $\tau_i$ amount of time allocated to $U_i$ for UL WIT in slot $i$, the average transmit power of $U_i$ is thus given by
   \begin{equation}\label{Eq_FD_TxPower_U_i}
      {{P_{{U_i}}} = \frac{{{\eta _i}{E_{{U_i}}}}}{{{\tau _i}}},\,\,\,\,\,\,i = 1,\,\, \cdots ,\,\,K,}
   \end{equation}
   where $0 < \eta_i < 1$ denotes the portion of the average harvested energy used for WIT by $U_i$ in steady state. In this paper, we assume that $\eta_i$'s are given constants. We further denote $x_{U,i}$ as the signal transmitted by $U_i$ during slot $i$. For each user, ${x_{U,i}}$ is assumed to be of zero mean and unit power, i.e., $\mathbb E[ {x_{U,i}}] = 0$ and $\mathbb E[ |{x_{U,i}}|^2] = 1$. The received signal at the H-AP in the $i$th slot is then expressed as
   \begin{equation}\label{Eq_FD_ReceivedSignal_AP}
      {{y_{A,i}} = \sqrt{P_{U_i}}{h_{U,i}}{x_{U,i}} + \sqrt{P_{i}}{g_{A}}{x_{A,i}} + {z_{A,i}}, \,\, i = 1, \, \cdots, \, K,}
   \end{equation}
   where ${y_{A,i}}$, $\sqrt{P_{i}}{g_{A}}{x_{A,i}}$, and ${z_{A,i}}$ denote the received signal, the SI due to simultaneously transmitted energy signal $x_{A,i}$, and the noise at the H-AP, respectively. It is assumed that ${z_{A,i}} \sim {\mathcal{CN}}\left( {0,\sigma^2} \right)$, $i = 1 \,\, \cdots, \,\, K$, where ${\mathcal{CN}}( {\nu ,\sigma^2})$ stands for a circularly symmetric complex Gaussian (CSCG) random variable with mean $\nu$ and variance $\sigma^2$.

   Without loss of generality, the effective loopback channel after analog domain SIC, $g_A$, can be expressed as
   \begin{equation}\label{Eq_FD_SI_After_Prop_Analog_Domain}
      {g_A = \sqrt{\varphi} \hat g_A,}
   \end{equation}
   where $\varphi$ and $\hat g_A$, respectively, denote the power and normalized complex coefficient of the loopback channel after analog domain SIC, with $\mathbb E [ {{\left| {{\hat g_{A}}} \right|}^2} ] = 1$. Even after analog domain SIC, however, the power of remaining SI can be still much larger than that of received signal in practice \cite{Kim}, i.e., $\left| h_{U,i} \right|^2 P_{U_i} \ll \varphi P_i$. This raises in the following considerations. First, the received signal at the H-AP given by (\ref{Eq_FD_ReceivedSignal_AP}) is distorted due to limited transmitter and receiver dynamic ranges due to large power of SI, which introduces additional noise whose power is proportional to transmit and/or receive power at the H-AP (see e.g., \cite{Day} and \cite{Kim}). For the purpose of exposition, we assume the same signal processing implemented at the H-AP as in \cite{Kim} with ideal automatic gain control (AGC) and infinite transmitter dynamic range. Thus, the additional receiver noise, denoted by $z_{Q,i}$, is merely due to finite receiver dynamic range and the resulting quantization error after ADC \cite{Kim}, where ${z_{Q,i}} \sim {\mathcal{CN}} ( 0, \beta \sigma_{Q,i}^2 )$ with $\beta \ll 1$ and $\sigma_{Q,i}^2$ is derived from (\ref{Eq_FD_ReceivedSignal_AP}) as
   \[
      {\sigma_{Q,i}^2 = \mathbb E \left[ \left| y_{A,i} \right|^2 \right] \,\,\,\,\,\,\,\,\,\,\,\,\,\,\,\,\,\,\,\,\,\,\,\,\,\,\,\,\,\,\,\,\,\,\,\,\,\,\,\,\,\,\,\,\,\,\,}
   \]
   \begin{equation}\label{Eq_QuantNoisePower}
      { = P_{U,i} \left| h_{U,i} \right|^2 + \varphi P_i + {\sigma ^2.}}
   \end{equation}

   Furthermore, although channel estimation can be made sufficiently accurate such that performance degradation due to the estimation errors in $h_{U,i}$ and $h_{D,i}$, $i = 1, \,\, \cdots, \,\, K$, is assumed to be negligible, performance degradation due to estimation error of $g_A$ cannot be neglected in general due to more dominant power of SI than that of received information signal at the H-AP. We thus denote $\hat g_A = \bar g_A + \sqrt{\varepsilon} \tilde g_A$ with $\bar g_A$ and $\sqrt{\varepsilon} \tilde g_A$ denoting the estimation of $\hat g_A$ and channel estimation error, respectively. It is assumed that $\tilde g_A \sim {\mathcal{CN}} ( 0, 1 )$ and $\varepsilon \ll 1$. Given $\bar g_A$, we can apply digital domain SIC by subtracting known SI as in \cite{Day}, i.e., $\sqrt{\varphi P_i} \bar g_A x_{A,i}$ from the received signal after ADC. After ADC and digital domain SIC, the received signal can be expressed as
   \[
      {{\bar y}_{A,i}} = {y_{A,i}} + {z_{Q,i}} - \sqrt {\varphi {P_1}} {{\bar g}_A}{x_{A,i}} \,\,\,\,\,\,\,\,\,\,\,\,\,\,\,\,\,\,\,\,\,\,\,\,\,\,\,\,\,\,\,\,\,\,\,\,\,\,\,\,\,\,\,\,\,\,\,\,\,\,\,\,\,\,\,\,\,\,\,\,\,\,
   \]
   \begin{equation}\label{Eq_FD_ReceivedSignal_AP_ADC_PSIC}
      {\,\,\,\,\,\,\,\, = \sqrt{P_{U_i}}{h_{U,i}}{x_{U,i}} + \sqrt{\varepsilon \varphi P_{i}}{\tilde g_{A}}{x_{A,i}} + z_{Q,i} + {z_{A,i}}, \,\, }
   \end{equation}
   \[
      i = 1, \,\, \cdots, \,\, K. \,\,\,\,\,\,\,\,\,\,\,\,\,\,\,\,\,\,\,\,\,\,\,\,\,\,\,\,\,\,\,\,\,\,\,\,\,\,\,\,\,\,\,\,\,\,\,\,\,\,\,\,\,\,\,\,\,\,\,\,\,\,\,\,\,\,\,\,\,\,\,\,
   \]
   It is worth noting that $\sigma_{Q,i}^2 \approx \varphi P_i$ in (\ref{Eq_QuantNoisePower}) since the power of SI is in general much larger than those of received information signal from $U_i$ and receiver noise, based on which the received signal-to-interference-plus-noise ratio (SINR) of $U_i$ after ADC and digital domain SIC can be expressed from (\ref{Eq_FD_ReceivedSignal_AP_ADC_PSIC}) as
   \begin{equation}\label{Eq_FD_SINR}
      {\rho_i =  {\frac{{{\theta _i}{H_i}}}{{ {\gamma {P_i} + {\sigma ^2}}}}\frac{1}{{{\tau _i}}}\sum\limits_{\scriptstyle j = 0 \hfill \atop
      \scriptstyle j \ne i \hfill}^K {{\tau _j}{P_j}} }, \,\,\,\, i = 1 , \,\, \cdots, \,\, K,}
   \end{equation}
   where $\theta_i = \eta_i \zeta_i$, $H_i = \left| h_{D,i} \right|^2 \left| h_{U,i} \right|^2$, and $\gamma P_i$ denotes the power of effective SI at the $i\,$th slot with
   \begin{equation}\label{Eq_Effective_SI}
      {\gamma = \varphi (\varepsilon + \beta).}
   \end{equation}
   From (\ref{Eq_FD_SINR}), the achievable rate of $U_i$ for the case of FD-WPCN in bits/second/Hz (bps/Hz) can be expressed as
   \[
      {{R_i^{(\rm{F})}}\left( {\boldsymbol{\tau} ,{\bf{P}}} \right) = {\tau _i}{\log _2}\left( 1 + \frac{\rho_i}{\Gamma} \right) \,\,\,\,\,\,\,\,\,\,\,\,\,\,\,\,\,\,\,\,\,\,\,\,\,\,\,\,\,\,\,\,\,\,\,\,\,\,\,\,\,\,\,\,\,\,\,\,\,\,\,\,\,\,\,\,\,\,\,\,\,\,\,\,\,\,\,\,\,\,\,\,\,\,\,\,\,\,\,\,\,\,\,\,\,\,\,\,\,\,\,\,\,\,\,\,\,\,\,\,\, }
   \]
   \begin{equation}\label{Eq_FD_AchievableRate}
      { \,\,\,\,\,\,\,\,\,\,\,\,\,\,\,\,\,\,\,\,\,\,\,\, = {\tau _i}{\log _2}\left( {1 + \frac{{{\theta _i}{H_i}}}{{\Gamma \left( {\gamma {P_i} + {\sigma ^2}} \right)}}\frac{1}{{{\tau _i}}}\sum\limits_{\scriptstyle j = 0 \hfill \atop
      \scriptstyle j \ne i \hfill}^K {{\tau _j}{P_j}} } \right),}
   \end{equation}
   where $\boldsymbol{\tau} = [\tau_{0}, \,\, \tau_{1}, \,\, \cdots \,\, \tau_{K}]$, ${\bf{P}} = [P_{0}, \,\, P_{1}, \,\, \cdots \,\, P_{K}]$, and $\Gamma$ represents the SINR gap from the additive white Gaussian noise (AWGN) channel capacity due to the practical modulation and coding scheme (MCS) used. In (\ref{Eq_FD_AchievableRate}), $\gamma P_i$ represents the residual SI power\footnote{Here, the residual SI includes the quantization noise whose power is also proportional to $P_i$.} due to finite receiver dynamic range and imperfect channel estimation. Furthermore, it is worth noting that switching transmit and receive antennas at the H-AP does not change the achievable rate in (\ref{Eq_FD_AchievableRate}) since ${R_i^{\rm{F}}}\left( {\tau ,{\bf{P}}} \right)$ depends only on $H_i$, which is the product of the UL and DL channel power gains.

   \subsection{HD-WPCN}

   \begin{figure}[!t]
       \centering
       \includegraphics[width=0.8\columnwidth]{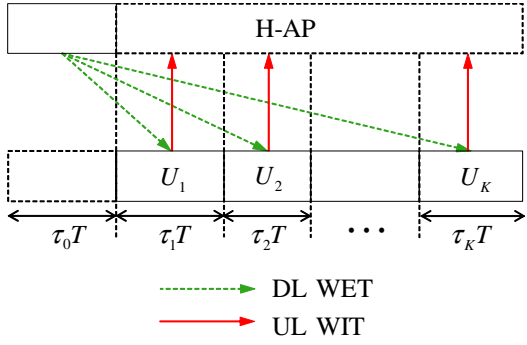}
       \caption{Orthogonal DL energy and UL information transmissions in HD-WPCN.}
       \label{Fig_HD_WPCN_FrameStructure}
   \end{figure}

   In the case of HD-WPCN, we assume that the \emph{harvest-then-transmit} protocol proposed in \cite{Ju} is employed to achieve orthogonal WET in DL and WIT of users in UL, as shown in Fig. \ref{Fig_HD_WPCN_FrameStructure}. Since neither H-AP nor users can transmit and receive signals at the same time, with the harvest-then-transmit protocol, energy is broadcast in DL during the $0\,$th slot only whereas the $i\,$th slot, $i = 1, \,\, \cdots, \,\, K$, is used for UL WIT only, in which users transmit their independent information to the H-AP by TDMA. It then follows that
   \begin{equation}\label{Eq_HD_SumTime}
      {\sum\limits_{i = 0}^{K} {{\tau _{i}}} \le 1.}
   \end{equation}
   Denote $P$ as the transmit power of the H-AP for DL WET in slot $i = 0$ in the case of HD-WPCN. The achievable rate of $U_i$ for UL WIT in bps/Hz is then given by \cite{Ju}
   \begin{equation}\label{Eq_HD_AchievableRate}
      { {R_i^{(\rm{H})}}\left( {\boldsymbol{\tau} , P} \right) =  {\tau _i}{\log _2}\left( {1 + \frac{{{\theta _i}{H_i}}}{{\Gamma {{\sigma ^2}} }} \frac{\tau_0 P}{{{\tau _i}}} } \right), \,\,\, i = 1, \,\, \cdots, \,\, K,}
   \end{equation}
   where ${\boldsymbol{\tau}} = [\tau_0 \,\, \tau_1, \,\, \cdots, \,\, \tau_K]$. Like the case of FD-WPCN, we have $P_{avg}$ and $P_{peak}$ as the average and peak transmit power constraints at the H-AP, respectively; thus we have $P = \min(P_{avg}/\tau_0,P_{peak})$.

\section{Optimal Time and Power Allocation in FD-WPCN}\label{Sec:FD_WPCN}
In this section, we study the joint time and power allocation in FD-WPCN to maximize the throughput. Specifically, from (\ref{Eq_FD_AchievableRate}), we aim to maximize the WSR of all users in UL WIT, which is formulated as the following optimization problem.
\[
   {({\rm{P1}}): \,\,\,\,\, \mathop {\max }\limits_{{\boldsymbol{\tau}}, {\bf{P}}} \,\,\,\,\,\sum\limits_{i = 1}^K {{\omega _i}{R_i^{(\rm{F})}}\left( {\boldsymbol{\tau} , \bf{P}} \right)} \,\,\,\,\,\,\,\,\,\,\,\,\,\,\,\,\,\,\,\,\,\,\,\,\,\,\,\,\,\,\,\,\,\,\,\,\,\, }
\]
\[
   { {\rm{s.t.}}\,\,\,\,\,\, (1), \,\, (2), \,\, {\rm{and}} \,\, (3), \,\,\,\,\,\,\,\,\,\,\, }
\]
\[
   { \,\,\,\,\,\,\,\,\,\,\,\,\,\,\,\,\,\,\,\,\,\,\,\,\,\,\,\,\,\,\,\,\,\,\,\,\,\,\,\,\,\,\,\,\,\,\,\,\,\,\,\,\,\,\, 0 \le \tau_i \le 1, \,\, P_i  \ge 0,  \,\,\,\, i = 0, \,\,1, \,\, \cdots \,\, K. }
\]

Note that $\omega_i$ is the given non-negative rate weight for $U_i$. Let ${\boldsymbol{\omega}} = [\omega_1, \,\, \cdots, \,\, \omega_K]$. By changing the values of $\omega_i$'s in (P1), we are able to characterize the maximum throughput of FD-WPCN with different rate trade-offs among the users. In the following, we first investigate time and power allocations to maximize the WSR in (P1) for an ideal FD-WPCN assuming perfect SIC, i.e., $\gamma = 0$ in (\ref{Eq_FD_AchievableRate}), and then study the general case with finite SI, i.e., $\gamma > 0$.

   \subsection{FD-WPCN with Perfect SIC}
   Without SI, the achievable rate of user $U_i$ in the $K$-user FD-WPCN is modified from (\ref{Eq_FD_AchievableRate}) as
   \begin{equation}\label{Eq_FD_AchievableRate_No_SI}
      {{R_i^{(\rm{F-NoSI})}}\left( {\boldsymbol{\tau} ,{\bf{P}}} \right) = {\tau _i}{\log _2}\left( {1 + \frac{{{\theta_i H_i}}}{{ {\Gamma \sigma ^2}}}\frac{1}{\tau_i}\sum\limits_{\scriptstyle j = 0 \hfill \atop
      \scriptstyle j \ne i \hfill}^K {{\tau _j}{P_j}} } \right).}
   \end{equation}

   Even with each user's achievable rate given by (\ref{Eq_FD_AchievableRate_No_SI}) assuming perfect SIC, problem (P1) is still non-convex due to the non-convexity shown in the average power constraint given by (\ref{Eq_FD_AvgPower}). Similarly as in \cite{Zhou_OFDM} and \cite{Orhan}, we change the variables as $E_i = \tau_i P_i$, $i = 0, \,\, 1, \,\, \cdots, \,\,K$, to make this problem more analytically tractable, where $E_i$ denotes the energy broadcast by the H-AP during the $i\,$th slot. From (\ref{Eq_FD_AchievableRate_No_SI}), the achievable rate of $U_i$ can then be expressed as
   \begin{equation}\label{Eq_FD_AchievableRate_No_SI_New}
      {\hat R_i^{(\rm{F-NoSI})}\left( { {\boldsymbol{\tau}} ,{\bf{E}}} \right) = {{\tau _i}{\log _2}\left( {1 + \alpha_i \frac{1}{\tau_i}\sum\limits_{\scriptstyle j = 0 \hfill \atop
      \scriptstyle j \ne i \hfill}^K {{E_j}} } \right)},}
   \end{equation}
   where ${\bf{E}} = [E_0, \,\, E_1, \,\, \cdots, \,\, E_K]$ and $\alpha_i = \frac{\theta_i H_i}{\Gamma \sigma^2}$. Accordingly, problem (P1) in the case of perfect SIC can be reformulated as
   \[
      { ({\rm{P2}}): \,\,\,\,\, \mathop {\max }\limits_{{\boldsymbol{\tau}}, {\bf{E}}} \,\,\,\,\, \sum\limits_{i = 1}^K {\omega _i} \hat R_i^{(\rm{F-NoSI})}\left( { {\boldsymbol{\tau}} ,{\bf{E}}} \right)  \,\,\,\,\,\,\,\,\,\,\,\,\,\,\,\,\,\,\,\,\,\,\,\,\,\,\,\,\,\,\,\,\,\,\,\,\,\,\,\,\,\,\,\,\,\,\,\, }
   \]
   \[
      {{\rm{s.t.}}\,\,\,\,\,\, \sum\limits_{i = 0}^K {{\tau _i}}  \le 1, \,\,\,\,\,\,\,\,\,\,\,\,\,\,\,\,\,\,\,\,\,\,\,\,\,\,\,\,\,\,\,\,\,\,\,\,\,\,\,\,\,\,\,\,\,\,\,\,\,\,\,\,\,\, }
   \]
   \begin{equation}\label{Eq_FD_SumEnergyConstraint}
       \sum\limits_{i = 0}^K {E_i}  \le {P_{avg}}, \,\,\,\,\,\,\,\,\,\,\,\,\,\,\,\,\,\,\,\,\,\,\,\,\,\,\,
   \end{equation}
   \begin{equation}\label{Eq_FD_PeakPowerConstraint_New}
      { \,\,\,\,\,\,\,\,\,\,\,\,\,\,\,\,\,\,\,\,\,\,\,\,\,\,\,\,\,\,\,\,\,\,\,\,\,\,\,\, E_i -{P_{peak}}\tau_i \le 0, \,\,\,\, i = 0, \,\, 1, \,\, \cdots \,\, K, }
   \end{equation}
   \[
      { \,\,\,\,\,\,\,\,\,\,\,\,\,\,\,\,\,\,\,\,\,\,\,\,\,\,\,\,\,\,\,\,\,\,\,\,\,\, 0 \le \tau_i \le 1, \,\, E_i \ge 0, \,\,\,\, i = 0, \,\, 1, \,\, \cdots \,\, K, }
   \]
   where (\ref{Eq_FD_SumEnergyConstraint}) and (\ref{Eq_FD_PeakPowerConstraint_New}) correspond to the original constraints in (\ref{Eq_FD_AvgPower}) and (\ref{Eq_FD_PeakPower}), respectively. By introducing new variables in $\bf{E}$, joint time and power allocation in problem (P1) is converted to joint time and energy allocation in problem (P2).

   It is worth noting that for any given $\boldsymbol{\tau}$, the objective function of problem (P2) is a monotonically increasing function of each individual $E_i$, $i = 0, \,\, 1, \,\, \cdots, \,\, K$, and thus the constraint in (\ref{Eq_FD_SumEnergyConstraint}) should hold with equality at the optimal energy allocation (otherwise, the objective function can be further increased by increasing some $E_i$'s). Therefore, the optimal time and energy allocation solution for (P2) can be equivalently obtained by solving the following problem:
   \[
      { ({\rm{P3}}): \,\,\,\,\,  \mathop {\max }\limits_{{\boldsymbol{\tau}}, {\bf{E}}} \,\,\,\,\, \sum\limits_{i = 1}^K {\omega _i} R_i^{(\rm{F-NoSI})}\left( { \boldsymbol{\tau} ,{\bf{E}}} \right) \,\,\,\,\,\,\,\,\,\,\,\,\,\,\,\,\,\,\,\,\,\,\, }
   \]
   \begin{equation}\label{Eq_FD_SumTime_New}
      {{\rm{s.t.}}\,\,\,\,\,\, \sum\limits_{i = 0}^K {{\tau _i}}  \le 1, \,\,\,\,\,\,\,\,\,\,\,\,\,\,\,\,\,\,\,\,\,\,\,\,\,\,\, }
   \end{equation}
   \begin{equation}\label{Eq_FD_SumEnergyConstraint_new}
       {\,\, \sum\limits_{i = 0}^K {E_i}  = {P_{avg}}, \,\,\,}
   \end{equation}
   \begin{equation}\label{Eq_FD_PeakPowerConstraint_New2}
      { \,\,\,\,\,\,\,\,\,\,\,\,\,\,\,\,\,\,\,\,\,\,\,\,\,\, E_i -{P_{peak}}\tau_i \le 0, \,\,\,\, i = 0, \,\, 1, \,\, \cdots \,\, K, }
   \end{equation}
   \begin{equation}\label{Eq_FD_PositiveConstraint}
      { \,\,\,\,\,\,\,\,\,\,\,\,\,\,\,\,\,\,\,\,\,\,\,\,\,\,\,\,\,\,\,\,\,\, 0 \le \tau_i \le 1, \,\, E_i \ge 0, \,\,\,\, i = 0, \,\, 1, \,\, \cdots \,\, K, }
   \end{equation}
   with $R_i^{(\rm{F-NoSI})}\left( { {\boldsymbol{\tau}} ,{\bf{E}}} \right)$ given by
   \begin{equation}\label{Eq_APP_FD_AchievableRate_No_SI_New}
      {R_i^{(\rm{F-NoSI})}\left( { {\boldsymbol{\tau}} ,{\bf{E}}} \right) = {{\tau _i}{\log _2}\left( {1 + \alpha_i \frac{1}{\tau_i} \left( P_{avg} - E_i \right)  } \right)}.}
   \end{equation}

   \emph{\underline{Lemma} 3.1:}
      ${R_i}^{\rm{F-NoSI}}\left( {\boldsymbol{\tau} ,{\bf{E}}} \right)$ is a jointly concave function of $\boldsymbol{\tau}$ and $\bf{E}$, $\forall i = 0, \,\, 1, \,\, \cdots, \,\, K$.
   \begin{proof}
      Please refer to Appendix \ref{App_Proof_FD_WSR_Concavity}.
   \end{proof}

   From Lemma 3.1, it follows that the objective function of (P3) is jointly concave over $\boldsymbol{\tau}$ and $\bf{E}$. Therefore, problem (P3) is a convex optimization problem together with the facts that the average power constraints in (\ref{Eq_FD_AvgPower}) of (P1), which is non-convex, is now transformed to a sum-energy constraint in (\ref{Eq_FD_SumEnergyConstraint_new}) of (P3), which is affine, and furthermore the constraint in (\ref{Eq_FD_PeakPowerConstraint_New2}) is an affine function of both $\boldsymbol{\tau}$ and $\bf{E}$. Problem $({\rm{P3}})$ can thus be solved by Lagrangian duality, shown as follows.

   From (\ref{Eq_FD_SumTime_New}), (\ref{Eq_FD_SumEnergyConstraint_new}), and (\ref{Eq_APP_FD_AchievableRate_No_SI_New}), Lagrangian of $({\rm{P3}})$ is given by
   \[
      \mathcal L\left( {\boldsymbol{\tau} ,{\bf{E}},\lambda ,\mu } \right) = \sum\limits_{i = 1}^K {\omega _i} R_i^{(\rm{F-NoSI})}\left( { \boldsymbol{\tau} ,{\bf{E}}} \right) \,\,\,\,\,\,\,\,\,\,\,\,\,\,\,\,\,\,\,\,\,\,\,\,\,\,\,\,\,\,\,\,\,\,\,\,\,\,\,\,\,\,\,\,\,\,\,\,\,\,
   \]
   \begin{equation}\label{Eq_Lagrangian}
      { \,\,\,\,\,\,\,\,\,\,\,\,\,\,\,\,\,\,\,\,\,\,\,\,\,\,\,\,\,\, - \lambda \left( {\sum\limits_{i = 0}^K {{\tau _i}}  - 1} \right) + \mu \left( {\sum\limits_{i = 0}^K {{E_i}}  - {P_{avg}}} \right),}
   \end{equation}
   with $\lambda \ge 0$ and $\mu$ denoting the Lagrange multipliers associated with the constraints in (\ref{Eq_FD_SumTime_New}) and (\ref{Eq_FD_SumEnergyConstraint_new}), respectively. The dual function of problem (P3) is then given by
   \begin{equation}\label{Eq_DualFunction}
      {\mathcal G\left( {\lambda ,\mu } \right) = \mathop {\max }\limits_{\left( {\boldsymbol{\tau} ,{\bf{E}}} \right) \in \mathcal D} \,\,\, \mathcal L\left( {\boldsymbol{\tau} ,{\bf{E}},\lambda ,\mu } \right),}
   \end{equation}
   where $\mathcal D$ is a feasible set of $(\boldsymbol{\tau}, \bf{E})$ specified by (\ref{Eq_FD_PeakPowerConstraint_New2}) and (\ref{Eq_FD_PositiveConstraint}). The dual problem of (P3) is thus given by $\mathop {\min }\limits_{\lambda  \ge 0,\mu} \,\,\, \mathcal G\left( {\lambda ,\mu } \right)$.

   \emph{\underline{Proposition} 3.1:}
      Given $\lambda \ge 0$, $\mu$, and strictly positive weights $\omega_i > 0$, $i = 1, \,\, \cdots, \,\, K$, the maximizer of $\mathcal L \left( {\boldsymbol{\tau} ,{\bf{E}},\lambda ,\mu } \right)$ in (\ref{Eq_Lagrangian}) is given by $\boldsymbol{\tau}^{\star} = [\tau_{0}^{\star}, \,\, \tau_{1}^{\star}, \,\, \cdots \,\, \tau_{K}^{\star}]$ and ${\bf{E}}^{\star} = [E_{0}^{\star}, \,\, E_{1}^{\star}, \,\, \cdots \,\, E_{K}^{\star}]$, where
      \begin{equation}\label{Eq_Prop_FD_Opt_tau_0}
         {\tau _0^{\star} = \left\{ {\begin{array}{*{20}{c}}
         {1}  \\
         {0}  \\
         \end{array}\begin{array}{*{20}{c}}
         {,\,\,{\rm{if}}\,\, \mu > 0 \,\, {\rm{and}} \,\, \left( -\lambda + \mu P_{peak}\right) > 0, }  \\
         {,\,\,\,\,\,\,\,\,\,\,\,\,\,\,\,\,\,\,\,\,\,\,\,\,\,\,\,\,\,\,\,\, {\rm{otherwise},} \,\,\,\,\,\,\,\,\,\,\,\,\,\,\,\,\,\,\,\,\,\,\,\,\,\,\,\,}  \\
         \end{array}} \right.}
      \end{equation}
      \begin{equation}\label{Eq_Prop_FD_Opt_E0}
         {E_0^{\star} = \left\{ {\begin{array}{*{20}{c}}
         {P_{peak}\tau_0^{\star}}  \\
         {0}  \\
         \end{array}\begin{array}{*{20}{c}}
         {,\,\,\, {\rm{if}}\,\, \mu > 0, \,}  \\
         {, {\rm{otherwise},} }  \\
         \end{array}} \right.}
      \end{equation}
      \begin{equation}\label{Eq_Prop_FD_Opt_tau_i}
         \tau_i^{\star} = {\min \left[ \left( {\frac{{{\alpha _i}}}{{z_i^{\star}}}\left( {{P_{avg}} - E_i^{\star}} \right)} \right)^+, \,\, 1 \right]}, \,\, i = 1, \, \cdots, \, K,
      \end{equation}
      \begin{equation}\label{Eq_Prop_FD_Opt_Ei}
         E_i^{\star} = \min \left[ {{{\left( {{P_{avg}} + \frac{{\tau _i^{\star}}}{{{\alpha _i}}} - \frac{{{\omega _i}\tau _i^{\star}}}{{{\mu}\ln 2}}} \right)}^ + }, \,\, {P_{peak}}\tau _i^{\star}} \right],
      \end{equation}
      \[
         \,\,\,\, i = 1, \,\, \cdots, \,\, K,
      \]
      with $\left(x\right)^+ \buildrel \Delta \over = \max \left( 0, x\right)$, and $z_i^*$ denoting the solution of $f\left( z_i \right) = \frac{\lambda^* \ln 2}{\omega_i}$ where
      \begin{equation}\label{Eq_Function_z}
         {f\left( z \right) \buildrel \Delta \over =   \ln \left( {1 + z} \right) - \frac{z}{{1 + z}}.}
      \end{equation}
   \begin{proof}
      Please refer to Appendix \ref{App_Proof_Prop_FD_NoSI_OptSolution}.
   \end{proof}

   According to Proposition 3.1, $\boldsymbol{\tau}^{\star}$ and ${\bf{E}}^{\star}$ under which $\mathcal G \left( \lambda, \mu \right)$ in (\ref{Eq_DualFunction}) is achived can be attained as follows. We first obtain $[\tau_1^{\star}, \,\, \cdots, \,\, \tau_K^{\star}]$ and $[E_1^{\star}, \,\, \cdots, \,\, E_K^{\star}]$ by iteratively optimizing between $[\tau_1, \,\, \cdots, \,\, \tau_K]$ and $[E_1, \,\, \cdots, \,\, E_K]$ using (\ref{Eq_Prop_FD_Opt_tau_i}) and (\ref{Eq_Prop_FD_Opt_Ei}), respectively, with one of them being fixed at one time until they both converge. We then compute $\tau_0^{\star}$ and $E_0^{\star}$ using (\ref{Eq_Prop_FD_Opt_tau_0}) and (\ref{Eq_Prop_FD_Opt_E0}), respectively. With $\mathcal G \left( \lambda, \mu \right)$ obtained for each given pair of $\lambda$ and $\mu$, the optimal dual variables $\lambda^*$ and $\mu^*$ minimizing $\mathcal G \left( \lambda, \mu \right)$ can then be efficiently found by sub-gradient based algorithms, e.g., the ellipsoid method \cite{LectureNote}, with the sub-gradient of $\mathcal G \left( \lambda, \mu \right)$ given by $\boldsymbol{\nu} = [\nu_{\lambda}, \,\, \nu_{\mu}]$, where
   \begin{equation}\label{Eq_Subgradient_Lambda}
      \nu_{\lambda} = \sum\limits_{i=0}^K \tau_i^{\star} - 1,
   \end{equation}
   \begin{equation}\label{Eq_Subgradient_Mu}
      \nu_{\mu} = P_{avg} - \sum\limits_{i=0}^K E_i^{\star}.
   \end{equation}

   Denote the optimal time and energy allocation solution for $({\rm{P3}})$, and equivalently for problem (P2), as $\boldsymbol{\tau}^{*} = [\tau_{0}^{*}, \,\, \tau_{1}^{*}, \,\, \cdots \,\, \tau_{K}^{*}]$ and ${\bf{E}}^{*} = [E_{0}^{*}, \,\, E_{1}^{*}, \,\, \cdots \,\, E_{K}^{*}]$, respectively. It is then worth noting that the objective function of (P3) is a monotonically increasing function of each individual $\tau_i$, $i = 1, \,\, \cdots, \,\, K$ for given $E_i$, $i = 1, \,\, \cdots, \,\, K$. Therefore, $\sum\limits_{i=0}^K \tau_i^* = 1$ should hold at the optimal $\boldsymbol{\tau}^*$ (otherwise, the objective function can be further increased by increasing some $\tau_i$'s). Furthermore, note that we obtain $[\tau_1^{*}, \,\, \cdots, \,\, \tau_K^{*}]$ and $[E_1^{*}, \,\, \cdots, \,\, E_K^{*}]$ at the optimal dual solution $\lambda^*$ and $\mu^*$. Therefore, we have $\tau_0^* = 1 - \sum\limits_{i=1}^K \tau_i^*$, as well as $E_0^* = P_{avg} - \sum\limits_{i=1}^K E_i^*$ to satisfy the constraint in (\ref{Eq_FD_SumEnergyConstraint_new}). Once ${\boldsymbol{\tau}}^*$ and ${\bf{E}}^*$ for (P3) (or equivalently for (P2)) are obtained, the optimal power allocation solution $\bf{P}^*$ for (P1) is obtained as $P_i^* = \frac{E_i^*}{\tau_i^*}$, $i = 0, \,\, 1, \,\, \cdots, \,\, K$, in the case of perfect SIC. To summarize, one algorithm to solve (P1) is given in Table \ref{Table_Algorithm_P2}.

   \begin{table}[!t]
   \renewcommand{\arraystretch}{1.3}
   \caption{Algorithm to solve (P1) with $\gamma = 0$.}
   \label{Table_Algorithm_P2} \centering
      \begin{tabular}{|p{3.2in}|}
      \hline
         1. Initialize $\lambda \ge 0$ and $\mu$.

         2. \textbf{Repeat}
            \begin{itemize}
               \item[1)] Initialize $\tau_i$ and $E_i$, $i = 1, \,\, \cdots, \,\, K$.

               \item[2)] \textbf{Repeat}

                  $\,\,\,\,$ i) Compute $[\tau_1, \,\, \cdots, \,\, \tau_K]$ by (\ref{Eq_Prop_FD_Opt_tau_i}).

                  $\,\,\,\,$ii) Compute $[E_1, \,\, \cdots, \,\, E_K]$ by (\ref{Eq_Prop_FD_Opt_Ei}).

               \item[3)] \textbf{Until} $[\tau_1, \,\, \cdots, \,\, \tau_K]$ and $[E_1, \,\, \cdots, \,\, E_K]$ both converge.
               \item[4)] Compute $\tau_0$ and $E_0$ by (\ref{Eq_Prop_FD_Opt_tau_0}) and (\ref{Eq_Prop_FD_Opt_E0}).

               \item[5)] Compute the sub-gradient of $\mathcal G \left( \lambda, \mu \right)$ by (\ref{Eq_Subgradient_Lambda}) and (\ref{Eq_Subgradient_Mu}).

               \item[6)] Update $\lambda$ and $\mu$ using the ellipsoid method.

            \end{itemize}

         3. \textbf{Until} $\lambda$ and $\mu$ converge to a predefined accuracy.

         4. Set $\tau_i^* = \tau_i$ and $E_i^* = E_i$, $i = 1, \,\, \cdots, \,\, K$.

         5. Obtain $\tau _0^* = 1 - \sum\nolimits_{i = 1}^K {\tau _i^*}$ and $E_0^* = P_{avg} - \sum\nolimits_{i = 1}^K {E_i^*}$.

         6. Set $P_i^* = \frac{E_i^*}{\tau_i^*}$, $i = 1, \,\, \cdots, \,\, K$.
               \\
      \hline
      \end{tabular}
   \end{table}

   The computation time of the algorithm given in Table \ref{Table_Algorithm_P2} is analyzed as follows. The time complexities of steps 2.1)-2.3) is $\mathcal O(K)$, while those of 2.4) and 2.5) are $\mathcal O(1)$ and $\mathcal O(K)$, respectively. Note that only two dual variables, $\lambda$ and $\mu$, are updated by the ellipsoid algorithm regardless of the number of users, $K$. The time complexity of step 2.6) is thus $\mathcal O(1)$ \cite{LectureNote}. Therefore, the total time complexity of the algorithm in Table \ref{Table_Algorithm_P2} is $\mathcal O(K)$.

   Next, to obtain more insight to the solution given in Proposition 3.1, we consider the special case of $P_{peak} = \infty$. From Proposition 3.1, $\boldsymbol{\tau}^*$ and ${\bf{E}}^*$ (or ${\bf{P}}^*$) for problem (P2) when $P_{peak} = \infty$ is given in the following corollary.

   \emph{\underline{Corollary} 3.1:}
      For problem (P2) with $P_{peak} = \infty$, the optimal time and power allocation solutions are given by
      \begin{equation}\label{Eq_Cor_OptEnergy_NoPPC}
         {\left( {\tau _i^*,P_i^*} \right) = \left\{ {\begin{array}{*{20}{c}}
         {\,\,\, \left( {0,{\infty}} \right)}  \\
         {\left( {\frac{{{\alpha _i}}}{{z_i^*}}{P_{avg}},0} \right)}  \\
         \end{array}\,\,\,\begin{array}{*{20}{c}}
         {,\,\,\,\,\,\,\,\,\, i = 0 \,\,\,\,\,\,\,}  \\
         {,\,\,\,{\rm{otherwise}},}  \\
         \end{array}} \right.}
      \end{equation}
      where $z_i^{*}$ denotes the solution of $f(z) = \frac{\lambda^{\star} \ln 2}{\omega_i}$ with $\lambda^{*} > 0$ and $f(z)$ is given by (\ref{Eq_Function_z}).
   \begin{proof}
      Please refer to Appendix \ref{App_Proof_Cor_FD_Max_WSR_Infinite_PPC}.
   \end{proof}

   Corollary 3.1 indicates that given rate-weight vector $\boldsymbol{\omega}$, the maximum WSR of FD-WPCN without SI as $P_{peak} \rightarrow \infty$ is achieved by choosing only the $0\,$th slot for DL WET, which has zero amount of time (i.e., $\tau_0^* \rightarrow 0$), and sending all available energy $P_{avg}$ (cf. (\ref{Eq_FD_SumEnergyConstraint})) in this zero-time slot (i.e., $P_0^* \rightarrow \infty$, $\tau_0^* P_0^* \rightarrow P_{avg}$). It is worth noting that in this case the WSR of the FD-WPCN is equivalent to that of the conventional $K$-user TDMA network for UL WIT only, where each user is equipped with a constant energy supply to provide equal transmit energy consumption at each block denoted by $P_{avg}$ and the UL channel for each user $i$ is given by $\alpha_i$, $i = 1, \,\, \cdots, \,\, K$. It is also worth noting that in this special case, the FD-WPCN is in fact a HD-WPCN with $P_{peak} \to \infty$ and $\tau_0 \to 0$. When the sum-rate is maximized with $\omega_i = 1$, $i = 1, \,\, \cdots, \,\, K$, it can be further shown that $\tau_0^* = 0$ and $\tau_i^* = {{\alpha _i}/\sum\limits_{j = 1}^K {{\alpha _j}} }$, $i = 1, \,\, \cdots, \,\, k$.

   Note that when $P_{peak} < \infty$, $\mu^*$ should be strictly positive since $0 \le E_i^* \le P_{avg}$ ($E_i^* \to -\infty$ if $\mu^* = 0$ and $E_i^* > P_{avg}$ if $\mu^* < 0$. Please see Appendix \ref{App_Proof_Prop_FD_NoSI_OptSolution} for details.). It is thus shown from (\ref{Eq_Prop_FD_Opt_E0}) that $E_0^* = P_{peak}\tau_0^*$. It thus follows that if $E_0^* = 0$, then $\tau_0^* = 0$ and thus $P_0^*=0$, and vice versa if $\tau_0^*=0$. For $i \ne 0$, $\tau_i^* = 0$ results in $E_i^* = 0$ as shown from (\ref{Eq_Prop_FD_Opt_Ei}), whereas $E_i^* = 0$ does not necessarily imply $\tau_i^* = 0$ as shown from (\ref{Eq_Prop_FD_Opt_tau_i}). This is because $E_i$, $i = 1, \,\, \cdots, \,\, K$, represents the amount of energy that the H-AP broadcasts to $U_j$'s, $j = 1, \,\, \cdots, \,\, K$, $j \ne i$, during the $i\,$th slot, while provided that $\tau_i^* >0$, $U_i$ can transmit information to the H-AP even when the H-AP does not broadcast any energy during the slot, i.e., $E_i = 0$. Therefore, even when $P_{peak} < \infty$, it is possible to allocate zero transmit power to a slot with $i \ne 0$ at the H-AP, i.e., $P_i^* = 0$. The only case where $E_i^* = 0$ for given $i \ne 0$ can yield $\tau_i^* = 0$ is when the corresponding $\omega_i = 0$ since $z_i^*$ should be zero for this case, according to (\ref{Eq_Function_z}).

   \begin{figure}[!t]
      \centering
      \includegraphics[width=1.0\columnwidth]{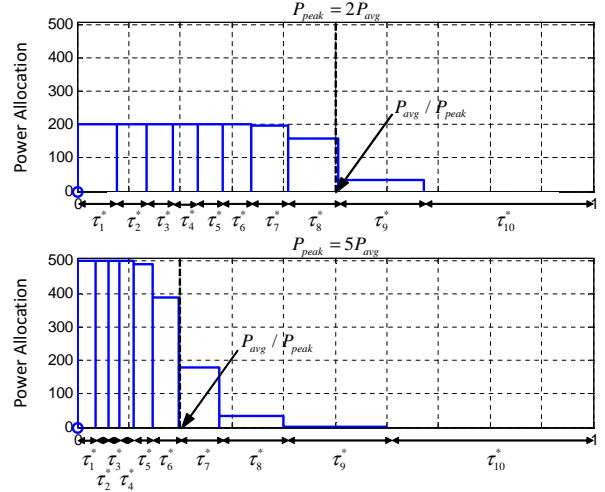}
      \caption{Time and power allocations to maximize sum-rate in FD-WPCN with perfect SIC for $K = 10$, $P_{peak} = 2P_{avg}$ and $P_{peak} = 5P_{avg}$.}
      \label{Fig_TimeAlloc_vs_PowerAlloc}
   \end{figure}

   It is also worth noting from Corollary 3.1 that in the case of $P_{peak} = \infty$, the algorithm given in Table \ref{Table_Algorithm_P2} can be similarly applied to update $\boldsymbol{\tau}$ and $\bf{E}$ (or equivalently $\bf{P}$) from any arbitrary point ($\boldsymbol{\tau}^{(0)}$,${\bf{P}}^{(0)}$) toward $(\boldsymbol{\tau}^{*}, {\bf{P}}^{*})$ given in (\ref{Eq_Cor_OptEnergy_NoPPC}). With $P_{peak} < \infty$, however, $P_i$, $i = 1, \,\, \cdots, \,\, K$, stops updating when it reaches $P_i^* = P_{peak}$, after which $(\tau_i, E_i)$ is updated such that $E_i = P_{peak} \tau_i$. Fig. \ref{Fig_TimeAlloc_vs_PowerAlloc} shows $\boldsymbol{\tau}^*$ and ${\bf{P}}^*$ to maximize the sum-rate in a SI-free FD-WPCN with $K = 10$, $P_{avg} = 100$, $\sigma^2 = 1$,\footnote{In Fig. \ref{Fig_TimeAlloc_vs_PowerAlloc}, $P_{avg}$ and $\sigma^2$ are normalized for the convenience of illustration. In fact, this corresponds to the case with $P_{avg} = 20$dBm in the practical simulation setup of Section \ref{Sec:SimulationResults}. Please refer to Section \ref{Sec:SimulationResults} for the detailed simulation setup.} $\Gamma = 1$, and $\theta_i = 1$, $\forall i$. We consider two different values of maximum peak power, $P_{peak} = 2P_{avg}$ or $5P_{avg}$. Furthermore, the DL and UL channels are assumed to be drawn from Rayleigh fading, i.e., $h_{D,\,i} \sim \mathcal{CN} (0,1)$ and $h_{U,\,i} \sim \mathcal{CN} (0,1)$, $\forall i$, and sorted in an ascending order of $i$ in terms of $\alpha_i$, i.e., $\alpha_1 < \alpha_2 < \cdots < \alpha_{10}$. For both cases with $P_{peak} = 2\,P_{avg}$ and $5\,P_{avg}$, it is observed that no time and power is allocated to the $0\,$th slot (dedicated power slot), i.e., $(\tau_0^*,P_0^*) = (0,0)$, to maximize the sum-rate. When $P_{peak} = 2P_{avg}$, it is observed that $P_1^* = \cdots = P_7^* = P_{peak}$, i.e., the H-AP broadcasts energy in DL with power of $P_{peak}$ during $1$st-$7$th slots in which the users with the weaker channel conditions transmit information in UL. Furthermore, $97\%$ of energy is broadcast from the H-AP during $1$st-$8$th slots, where $\sum\nolimits_{i = 1}^8 {\tau _i^*}  \approx \frac{{{P_{avg}}}}{{{P_{peak}}}} = 0.5$. The remaining time is allocated to the slots in which the users with better channel conditions transmit information in UL, and $U_i$ with larger $\alpha_i$ is allocated more time for UL WIT. In particular, it is also observed that $P_{10}^* = 0$, i.e., the H-AP does not broadcast energy in DL when the user with the best channel transmits information in UL in order to maximize the sum-rate. When $P_{peak} = 5P_{avg}$, similarly, $P_1^* = P_2^* = P_3^* = P_{peak}$ and the H-AP broadcasts $93\%$ of energy during the $1$st-$6$th slots where $\sum\nolimits_{i = 1}^6 {\tau _i^*}  = \frac{{{P_{avg}}}}{{{P_{peak}}}} = 0.2$. In addition, $\tau_7^* < \tau_8^* < \tau_9^* < \tau_{10}^*$ and $P_{10}^* = 0$. From the above observations, it can be inferred that to maximize the WSR, the H-AP does not broadcast energy in DL when the users with larger $\alpha_i$'s and/or $\omega_i$'s transmit information in UL, whereas it broadcasts energy only when the users with weaker channel conditions and/or smaller rate weights (priorities) transmit information.

\begin{figure}[!t]
      \centering
      \includegraphics[width=1.0\columnwidth]{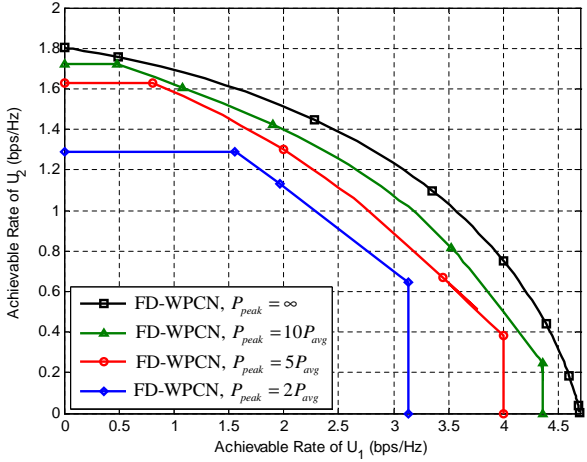}
      \caption{Achievable rate trade-offs in a 2-user FD-WPCN with perfect SIC.}
      \label{Fig_RateRegion_No_SI}
   \end{figure}

   Fig. \ref{Fig_RateRegion_No_SI} shows the achievable rate regions of a 2-user FD-WPCN with perfect SIC. It is assumed that $P_{avg} = 100$, $P_{peak} = 2P_{avg}$, $\sigma^2 = 1$, $\Gamma = 1$, and $\theta_i = 1$, $i = 1, \,\, \cdots, \,\, K$. The channels between the H-AP and two users $U_1$ and $U_2$ are set such that $H_1 = 0.249$ and $H_2 = 0.025$. We consider four different peak power values, $P_{peak} = 2P_{avg}$, $5P_{avg}$, $10P_{avg}$, and $\infty$. When $P_{peak} = \infty$, it is observed that the maximum achievable rate of $U_i$ is $\log_2 (1 + \alpha_i P_{avg})$ and the rate region of FD-WPCN in this case is equivalent to that of $K$-user TDMA network for WIT only, which is obtained with (\ref{Eq_Cor_OptEnergy_NoPPC}). With finite $P_{peak}$, the achievable rate region is observed to be smaller as $P_{peak}$ decreases.

   \subsection{FD-WPCN with Finite SI}\label{Sec:FD_WPCN_SI}
   In the case of imperfect SIC with $\gamma > 0$, problem (P1) is in general non-convex, and thus cannot be solved optimally in an efficient way. However, we can find one locally optimal solution for (P1) by setting $({\boldsymbol{\tau}}^*,{\bf{P}}^*)$ for the case without SI from Proposition 3.1 as an initial point, and then updating ${\boldsymbol{\tau}}$ and ${\bf{P}}$ iteratively, shown as follows. Denote ${\boldsymbol{\tau}}^{(k)}$ and ${\bf{P}}^{(k)}$ as the time and power allocation obtained at the $k\,$th iteration, respectively. We then have $({\boldsymbol{\tau}}^{(0)},{\bf{P}}^{(0)}) = ({\boldsymbol{\tau}}^*,{\bf{P}}^*)$. At the $k$-th iteration, we first update $\boldsymbol{\tau}$ to obtain ${\boldsymbol{\tau}}^{(k)}$ using ${\bf{P}}^{(k-1)}$. Then, $\bf{P}$ is updated to obtain ${\bf{P}}^{(k)}$ using $\boldsymbol{\tau}^{(k)}$ and ${\bf{P}}^{(k-1)}$. The above iteration is repeated until the WSR cannot be further improved for problem (P1).

   For simplifying the above updates, $R_i^{({\rm{F}})}\left( {{\boldsymbol{\tau}},{{\bf{P}}}} \right)$ given by (\ref{Eq_FD_AchievableRate}) is modified as
   \begin{equation}\label{Eq_FD_AchievableRate_SI}
      {R_i^{({\rm{F-SI}})}\left( {\tau_i , P_i} \right) = {\tau _i}{\log _2}\left( {1 + \frac{{{{\theta }_i}{H_i}}}{\Gamma \left({{\gamma P_i} + {{\sigma }^2}}\right)}\frac{{{P_{avg}} - {\tau_i P_i}}}{\tau_i}} \right),}
   \end{equation}
   by assuming that the average power constraint in (\ref{Eq_FD_AvgPower}) holds with equality.\footnote{The average power constraint in (\ref{Eq_FD_AvgPower}) may not hold with equality at the optimal time and power allocations in general for (P1), although this is usually desirable in practice since there is no energy wasted at the H-AP. We make this assumption here in order to have the rate function $R_i^{({\rm{F-SI}})}$ in (\ref{Eq_FD_AchievableRate_SI}) only dependent on $\tau_i$ and $P_i$, but no more on other $\tau_j$'s and $P_j$'s, $\forall j \ne i$, as in (13).} From (\ref{Eq_FD_AchievableRate_SI}), ${\boldsymbol{\tau}}^{(k)}$ at the $k\,$th iteration can be obtained as the solution of the following problem.
   \[
      {\,\,\,\,\,\,\,\,\,\,\,\,\,\,\, ({\rm{P4}}): \,\,\,\,\, \mathop {\max }\limits_{{\boldsymbol{\tau}}} \,\,\,\,\,\sum\limits_{i = 1}^K {{\omega _i} {R_i^{({\rm{F - SI}})}\left( {{\tau _i},P_i^{(k-1)}} \right)} }}
   \]
   \begin{equation}\label{Eq_FD_SumTime_NewNew}
      {{\rm{s.t.}}\,\,\,\,\,\, \sum\limits_{i = 0}^K {\tau_i}  \le {1},  }
   \end{equation}
   \begin{equation}\label{Eq_FD_AvgPower_New}
      { \,\,\,\,\,\,\,\,\,\,\,\,\,\,\,\,\,\,\,\,\,\,\,\,\,\,\,\,\,\,\,\,\,\,\,\,\,\,\,\,\, \sum\limits_{i = 0}^K {P_i^{(k-1)} \tau_i}  = {P_{avg}}, }
   \end{equation}
   \[
      { \,\,\,\,\,\,\,\,\,\,\,\,\,\,\,\,\,\,\,\,\,\,\,\,\,\,\,\,\,\,\,\,\,\,\,\,\,\,\,\,\,\,\,\,\,\,\, \tau_i  \ge 0, \,\,\,\, i = 0, \,\, \cdots \,\, K. }
   \]

   \emph{\underline{Lemma} 3.2:}
      Given ${\bf{P}} = {\bf{P}}^{(k-1)}$,
      ${R_i^{(\rm{F-SI})}} ( {\tau_i , P_i^{(k-1)}} )$ given in (\ref{Eq_FD_AchievableRate_SI}) is a concave function of $\tau_i$, $\forall i = 1 \,\, \cdots, \,\, K$.
   \begin{proof}
      Please refer to Appendix \ref{App_Proof_FD_SI_Concavity}.
   \end{proof}

   From Lemma 3.2, it follows that problem (P4) is convex. Therefore, problem (P4) can be solved by Lagrangian duality. From (\ref{Eq_FD_AchievableRate_SI})-(\ref{Eq_FD_AvgPower_New}), the Lagrangian of (P4) is given by
   \begin{equation}\label{Eq_Lagrangian_SI}
      {\mathcal L^{(\rm{F-SI})}\left( {\boldsymbol{\tau} , \lambda ,\mu } \right) = \sum\limits_{i = 1}^K {{\omega _i}R_i^{({\rm{F-SI}})}\left( { \tau_i ,{P_i^{(k-1)}}} \right)} \,\,\,\,\,\,\,\,\,\,\,\,\,\,\,\,\,\,\,\,\,\,\,\,\,\,\,\,\,\,\,\,\,\,\,\,\,\,\,\,\,\,\,\,\,\,\,\,\,\,\,\,\,\,\,\,\,\,\,\,\,\,\,\, }
   \end{equation}
   \[
      \,\,\,\,\,\,\,\,\,\,\,\,\,\,\,\,\,\,\,\,\,\,\,\,\,\, - \lambda \left( {\sum\limits_{i = 0}^K {{\tau _i}}  - 1} \right) - \mu \left( {\sum\limits_{i = 0}^K {{P_i^{(k-1)} \tau_i}}  - {P_{avg}}} \right),
   \]
   where $\lambda \ge 0$ and $\mu$ denote the Lagrange multipliers associated with the constraints in (\ref{Eq_FD_SumTime_NewNew}) and (\ref{Eq_FD_AvgPower_New}), respectively. The dual function of problem (P4) is then given by $\mathcal G^{(\rm{F-SI})}\left( {\lambda ,\mu } \right) = \mathop {\max }\limits_{ {\boldsymbol{\tau} } \ge 0} \,\,\, \mathcal L^{(\rm{F-SI})}\left( {\boldsymbol{\tau} , \lambda ,\mu } \right)$, and the dual problem of (P4) is thus given by $\mathop {\min }\limits_{\lambda  \ge 0, \, \mu} \,\,\, \mathcal G^{(\rm{F-SI})}\left( {\lambda ,\mu } \right)$.

   \emph{\underline{Proposition} 3.2:}
      Given ${\bf{P}}^{(k-1)}$, $\lambda$ and $\mu$ where $\lambda + \mu P_{avg} \ge 0$, the maximizer of $\mathcal L^{(\rm{F-SI})}\left( {\boldsymbol{\tau} , \lambda ,\mu } \right)$ is given by ${\boldsymbol{\bar \tau}} = [\bar \tau_1, \,\, \cdots, \,\, \bar \tau_K]$, where
      \begin{equation}\label{Eq_Prop_FD_SI_OptTime}
         {{\bar \tau _i} = \left\{ {\begin{array}{*{20}{c}}
         0  \\
         {\frac{{{C_i}{P_{avg}}}}{{z_i^* + {C_i}P_i^{(k - 1)}}}}  \\
         \end{array} \begin{array}{*{20}{c}}
         {, \,\,\,\,\,\,\,\,\,\,\,\,\, i = 0 \,\,\,\,\,\,\,\,\,\,\,\,\,\, }  \\
         {, \,\,\, i = 1, \,\, \cdots , \,\, K,}  \\
         \end{array}} \right.}
      \end{equation}
      where $C_i = \theta_i H_i / \left( \Gamma (\gamma P_i^{(k-1)} + \sigma^2) \right)$ and $z_i^*$ is the solution of $\bar f(z_i) = (\lambda + P_i^{(k)}\mu ){\ln 2}/\omega_i$ with $\bar f(z_i)$ defined as
      \begin{equation}\label{Eq_Function_bar_z}
         {\bar f\left( z \right) \buildrel \Delta \over =   \ln \left( {1 + z} \right) - \frac{z}{{1 + z}} - \frac{C_i P_i^{(k-1)}}{1+z}.}
      \end{equation}
   \begin{proof}
      Please refer to Appendix \ref{App_Proof_Prop_FD_SI_OptTime}.
   \end{proof}

   It can be shown that $\bar f\left( z \right)$ is a monotonically increasing function of $z$ since $\frac{\partial }{{\partial z}} \bar f\left( z \right) = \frac{z + C_i P_i^{(k-1)}}{(1+z)^2} > 0$. After obtaining $\mathcal G^{(\rm{F-SI})}\left( {\lambda ,\mu } \right)$ with given $\lambda$ and $\mu$, the minimization of $\mathcal G^{(\rm{F-SI})}\left( {\lambda ,\mu } \right)$ over $\lambda$ and $\mu$ can be efficiently solved by the ellipsoid method \cite{LectureNote}, with the sub-gradient of $\mathcal G^{(\rm{F-SI})}\left( {\lambda ,\mu } \right)$, denoted as ${\boldsymbol{\upsilon}} = {[ {\upsilon _{\lambda}, \,\upsilon _{\mu}} ]^T}$, given by
   \begin{equation}\label{Eq_FD_Subgradient_lambda}
      {\upsilon _{\lambda} = \sum\limits_{i = 0}^K {{{\bar \tau }_i}}  - 1,}
   \end{equation}
   \begin{equation}\label{Eq_FD_Subgradient_mu}
      {\upsilon _{\mu} = \sum\limits_{i = 0}^K {P_i^{(k-1)}{{\bar \tau }_i}}  - {P_{avg}},}
   \end{equation}
   where ${{\bar \tau }_i}$, $i = 0, \,\, \cdots, \,\, K$, are given by (\ref{Eq_Prop_FD_SI_OptTime}). The optimal solution of (P4), denoted by $\boldsymbol{\bar \tau}^*$, is then obtained corresponding to the optimal dual solution of $\lambda^*$ and $\mu^*$ after the ellipsoid method converges, and finally we have ${\boldsymbol{\tau}}^{(k)} = \boldsymbol{\bar \tau}^*$ for given ${\bf{P}}^{(k-1)}$.

   Once ${\boldsymbol{\tau}}^{(k)}$ is obtained, ${\bf{P}}^{(k)}$ at the $k\,$th iteration can be attained using the obtained ${\boldsymbol{\tau}}^{(k)}$. Denote ${\bf{P}}^{(k)}$ at the $k\,$th iteration as ${\bf{P}}^{(k)} = [P_0^{(k)} \,\,\, {\bar{\bf{P}}}^{(k)}]$, where ${\bar{\bf{P}}}^{(k)} = [P_1^{(k)}, \,\, \cdots, \,\, P_K^{(k)}]$. Since $\tau_0^{(k)} = 0$, $k \ge 1$, as shown in (\ref{Eq_Prop_FD_SI_OptTime}), we can thus take any value for $P_0^{(k)}$ such that $0 \le P_0^{(k)} \le P_{peak}$. Furthermore, ${\bar{\bf{P}}}^{(k)}$ can be obtained by applying the gradient projection method based on (\ref{Eq_FD_AchievableRate_SI}), as follows:
   \begin{equation}\label{Eq_FD_GP_K_1}
      {{{{\bf{\tilde P}}}^{(k)}} = {\mathcal P_{\mathcal E}}\left( {{\bar{\bf{P}}^{(k-1)}} + {s^{(k)}}\nabla \mathcal W_{\rm{F-SI}}\left( {{\boldsymbol{\tau} ^{(k)}}, {\bar{\bf{P}}^{(k-1)}}} \right)} \right),}
   \end{equation}
   \begin{equation}\label{Eq_FD_GP_K}
      {{\bar{\bf{P}}^{(k)}} = {\bar{\bf{P}}^{(k-1)}} + {\delta ^{(k)}}\left( {{{{\bf{\tilde P}}}^{(k)}} - {\bar{\bf{P}}^{(k-1)}}} \right),}
   \end{equation}
   where $\delta^{(k)} \in \left( {0,1} \right]$ and $s^{(k)}$ are both small step sizes. In addition, $\nabla \mathcal W_{\rm{F-SI}}\left( {{\boldsymbol{\tau} ^{(k)}}, {\bar{\bf{P}}^{(k-1)}}} \right) = [q_1^{(k)}, \,\, \cdots, \,\, q_K^{(k)}]^T$ denotes the gradient of $\mathcal W_{\rm{F-SI}} ( {{\boldsymbol{\tau} ^{(k)}}, {\bar{\bf{P}}^{(k-1)}}} ) \buildrel \Delta \over = \sum\nolimits_{i = 1}^K {R_i^{({\rm{F - SI}})} ( {\tau _i^{(k)},P_i^{(k - 1)}} )}$ with $q_i^{(k)}$, $i = 1, \,\, \cdots, \,\, K$, given by
   \begin{equation}\label{Eq_FD_Gradient}
      {q_i^{(k)} = \,\,\,\,\,\,\,\,\,\,\,\,\,\,\,\,\,\,\,\,\,\,\,\,\,\,\,\,\,\,\,\,\,\,\,\,\,\,\,\,\,\,\,\,\,\,\,\,\,\,\,\,\,\,\,\,\,\,\,\,\,\,\,\,\,\,\,\,\,\,\,\,\,\,\,\,\,\,\,\,\,\,\,\,\,\,\,\,\,\,\,\,\,\,\,\,\,\,\,\,\,\,\,\,\,\,\,\,\,\,\,\,\,\,\,\,\,\,\,\,\,\,\,\,\,\,\,\,\,\,\,\,\,\,\,\,\,\,\,\, }
   \end{equation}
   \[
      - \frac{{{\omega _i}\tau _i^{(k)}}}{{\ln 2}} \frac{\theta_i H_i \left( \tau_i^{(k)} + \gamma \left( P_{avg} - \tau_i^{(k)} P_i^{(k-1)} \right) \right)}{\Gamma \left( \gamma P_i^{(k-1)}  + \sigma^2 \right) \tau_i^{(k)}  + \theta_i H_i \left( P_{avg} - \tau_i^{(k)} P_i^{(k-1)} \right)}.
   \]
   Furthermore, $\mathcal E$ denotes the feasible set of ${\bf{P}}$ given $\boldsymbol{\tau}^{(k)}$ and $P_0^{(k)}$, defined by
   \begin{equation}\label{Eq_FD_FeasibleSet_Tau}
      {\mathcal E = \left\{ {{\bf{P}}\left| {\sum\limits_{i = 1}^K {\tau _i^{(k)}{P_i} = {P_{avg}}, \, 0 \le {P_i} \le {P_{peak}},i = 1, \cdots , K} } \right.} \right\}. }
   \end{equation}
   Finally, in (\ref{Eq_FD_GP_K_1}) $\mathcal P_{\mathcal E}({\bf{x}})$ denotes the operation of projection of $\bf{x}$ onto $\mathcal E$. To summarize, one algorithm to solve problem (P1) with finite $\gamma$ is given in Table \ref{Table_Algorithm_P3}.

   \begin{table}[!t]
   \renewcommand{\arraystretch}{1.3}
   \caption{Algorithm to solve (P1) with finite $\gamma$.}
   \label{Table_Algorithm_P3} \centering
      \begin{tabular}{|p{3.2in}|}
      \hline
         1.  Initialize $k=0$, $({\boldsymbol{\tau}}^{(0)}, {\bf{P}}^{(0)}) = ({\boldsymbol{\tau}}^*, {\bf{P}}^*)$ obtained by solving (P1).

         2. \textbf{Repeat}
            \begin{itemize}
               \item[1)] Set $k \leftarrow k+1$.

               \item[2)] Given $\lambda$, $\mu$, and ${\bf{P}}^{(k-1)}$, solve (P4) by Proposition 3.2.

               \item[3)] Update $\lambda$ and $\mu$ using the ellipsoid method and sub-gradient of $\mathcal G^{(\rm{F-SI})}(\lambda,\mu)$ given by (\ref{Eq_FD_Subgradient_lambda}) and (\ref{Eq_FD_Subgradient_mu}), and obtain $\boldsymbol{\bar \tau}^*$.

               \item[4)] Set $\boldsymbol{\tau}^{(k)} \leftarrow \boldsymbol{\bar \tau}^*$.

               \item[5)] Obtain ${\bf{P}}^{(k)}$ using ${\bf{P}}^{(k-1)}$, ${\boldsymbol{\tau}}^{(k)}$, (\ref{Eq_FD_AchievableRate_SI}), and (\ref{Eq_FD_GP_K_1})-(\ref{Eq_FD_FeasibleSet_Tau}).
            \end{itemize}

         3. \textbf{Until} $\mathcal W_{\rm{F-SI}} ({\boldsymbol{\tau} ^{(k-1)}}, {{\bf{P}}^{(k-1)}}) \ge \mathcal W_{\rm{F-SI}} ({\boldsymbol{\tau} ^{(k)}}, {{\bf{P}}^{(k)}})$.
               \\
      \hline
      \end{tabular}
   \end{table}

   At each iteration of the algorithm given in Table \ref{Table_Algorithm_P3}, the computational complexity of step 2.2) is $\mathcal O(K)$. Similarly to the case with perfect SIC in the previous subsection, $\mathcal O(K)$ computations are required for step 2.3). In step 2.5), $\mathcal O(K)$, $\mathcal O(K^2)$, and $\mathcal O(K)$ computations are required for computing $\nabla \mathcal W_{\rm{F-SI}}\left( {{\boldsymbol{\tau} ^{(k)}}, {\bar{\bf{P}}^{(k-1)}}} \right)$ in (\ref{Eq_FD_Gradient}), ${{{\bf{\tilde P}}}^{(k)}}$ in (\ref{Eq_FD_GP_K_1}), and ${\bar{\bf{P}}^{(k)}}$ in (\ref{Eq_FD_GP_K}), respectively. Therefore, the total time complexity of the algorithm in Table \ref{Table_Algorithm_P3} is $\mathcal O(K^2)$.

   \emph{\underline{Remark} 3.1:}
      With $P_{peak} \to \infty$, we can choose $P_0^* \rightarrow \infty$ such that $\tau_0^* P_0^* \rightarrow P_{avg}$, resulting in $P_i = 0$, $i = 1, \,\, \cdots, \,\, K$. In this special case, the FD-WPCN with finite SI is in fact equivalent to that in the previous ideal case of perfect SIC with $P_{peak} \to \infty$.

\section{Optimal Time and Power Allocation in HD-WPCN}\label{Sec:HD_WPCN}
In this section, we study the optimal time and power allocation in HD-WPCN to maximize the WSR. Given $P = \min(P_{avg}/\tau_0, P_{peak})$, from (\ref{Eq_HD_AchievableRate}) the WSR maximization problem for HD-WPCN is formulated as
\[
   {({\rm{P5}}): \,\,\,\,\, \mathop {\max }\limits_{{\boldsymbol{\tau}}, P} \,\,\,\,\,\sum\limits_{i = 1}^K {{\omega _i}{R_i^{(\rm{H})}}\left( {\boldsymbol{\tau} , P} \right)} \,\,\,\,\,\,\,\, }
\]
\begin{equation}\label{Eq_HD_SumTimeConstraint}
   {  {\rm{s.t.}}\,\,\,\,\,\, \sum\limits_{i = 0}^{K} {{\tau _{i}}} \le 1, }
\end{equation}
\begin{equation}\label{Eq_HD_AvgPower}
   { \,\,\,\,\,\,\,\,\,\,\,\,\,\,\,\,\,\,\,\,\, \tau_0 P \le P_{avg},}
\end{equation}
\begin{equation}\label{Eq_HD_PeakPower}
   { \,\,\,\,\,\,\,\,\,\,\,\,\,\,\,\,\,\, P \le P_{peak}.}
\end{equation}
\[
   { \,\,\,\,\,\,\,\,\,\,\,\,\,\,\,\,\,\,\,\,\,\,\,\,\,\,\,\,\,\,\,\,\,\,\,\,\,\,\,\,\,\,\,\,\,\,\,\,\,\,\,\,\,\,\,\,\,\,\,\,\,\,\,\,\,\,\,\,\,\, P  \ge 0, \,\, \tau_i  \ge 0, \,\,\,\, i = 0, \,\,1, \,\, \cdots \,\, K. }
\]

Problem (P4) is non-convex in general due to the non-concave objective function defined in (\ref{Eq_HD_AchievableRate}) and non-convex average power constraint in (\ref{Eq_HD_AvgPower}). To solve problem (P4), we first consider the following WSR maximization problem for HD-WPCN when the transmit power of the H-AP is fixed as $P = P_{peak}$.
\[
   { \,\,\,\,\,\,\,\,\,\,\,\, ({\rm{P6}}): \,\,\,\,\, \mathop {\max }\limits_{{\boldsymbol{\tau}}} \,\,\,\,\, \sum\limits_{i = 1}^K {\omega _i}{{\tau _i}{\log _2}\left( {1 + \alpha_i P_{peak} \frac{\tau_0}{\tau_i}} \right)} }
\]
\[
   {{\rm{s.t.}}\,\,\,\,\,\, \sum\limits_{i = 0}^K {{\tau _i}}  \le 1, \,\,\,\,\,\,\,\,\,\,\,\, }
\]
\[
   { \,\,\,\,\,\,\,\,\,\,\,\,\,\,\,\,\,\,\,\,\,\,\,\,\,\,\,\,\,\,\,\,\,\,\,\,\,\,\,\,\,\, \tau_i  \ge 0, \,\,\,\, i = 0, \,\,1, \,\, \cdots \,\, K, }
\]
with $\alpha_i$ defined in (\ref{Eq_FD_AchievableRate_No_SI_New}). As shown in \cite{Ju}, (P5) is convex and its optimal time allocation solution, denoted by ${\boldsymbol{\tau}}^{\star} = [\tau_0^{\star}, \,\, \tau_1^{\star}, \,\, \cdots, \,\, \tau_K^{\star}]$, is given in the following lemma.

\emph{\underline{Lemma} 4.1:}
   For problem (P6), the optimal solution is given by
   \begin{equation}\label{Eq_Lemma_HD_tau_0}
      {{\tau} _0^{\star} = \frac{1}{{1 + P_{peak} \sum\nolimits_{j = 1}^K {\left( {{\alpha _j}/{z_j}} \right)} }},}
   \end{equation}
   \begin{equation}\label{Eq_Lemma_HD_tau_i}
      {{\tau} _i^{\star} = \frac{P_{peak} \, \alpha_i / {{z}_i^{\star}} }{{1 + P_{peak} \sum\nolimits_{j = 1}^K {\left( {{\alpha _j}/{z_j}} \right)} }}, \,\,\, i = 1, \,\, \cdots, \,\, K,}
   \end{equation}
   where ${z}_i^{\star}$, $i = 1, \,\, \cdots, \,\, K$, is the solution of the following equations:
   \begin{equation}\label{Eq_Lemma_HD_FunctionZ}
      {\ln \left( {1 + {z_i}} \right) - \frac{{{z_i}}}{{1 + {z_i}}} = \frac{\nu^{\star}}{{{\omega _i}}}\ln 2, }
   \end{equation}
   \begin{equation}\label{Eq_Lemma_HD_KKT}
      {\sum\limits_{i = 1}^K {\frac{{{\omega _i}{\alpha_i}}}{{1 + {z_i}}}}  = \frac{{{{\nu}^{\star}}}\ln 2}{P_{peak}},}
   \end{equation}
   with ${\nu}^{\star} > 0$ being a constant under which $\sum_{i=0}^K \tau_i^{\star} = 1$.
\begin{proof}
   Please refer to \cite{Ju}.
\end{proof}

With ${\boldsymbol{\tau}}^{\star}$ attained from Lemma 4.1, we can obtain the optimal power and time allocation solution for (P5), denoted by $P^*$ and ${\boldsymbol{\tau}}^* = [\tau_0^*, \,\, \tau_1^*, \,\, \cdots, \tau_K^*]$, as given in the following proposition.

\emph{\underline{Proposition} 4.1:}
   The optimal transmit power solution for (P5) is $P^* = P_{peak}$. In addition, the optimal time allocation solution for (P5) is given by
   \begin{equation}\label{Eq_Prop_HD_Opt_DL_Time}
      {\tau _0^* = \left\{ {\begin{array}{*{20}{c}}
      {\tau _0^{\star}}  \\
      {\frac{{{P_{avg}}}}{{{P_{peak}}}}}  \\
      \end{array}\begin{array}{*{20}{c}}
      {,\,\,\,{\rm{if}} \,\,\, 0 < \tau _0^{\star} \le \frac{{{P_{avg}}}}{{{P_{peak}}}}}  \\
      {, \,\,\,\,\,\,\,\,\,\,\,\, {\rm{otherwise}}, \,\,\,\,\,\,\,\,\, }  \\
      \end{array}} \right.}
   \end{equation}
   \begin{equation}\label{Eq_Prop_HD_Opt_UL_time}
      {\tau _i^* = \left\{ {\begin{array}{*{20}{c}}
      {\tau _i^{\star}}  \\
      {\frac{{{P_{avg}}{\alpha _i}}}{{{z_i^*}}}}  \\
      \end{array}\begin{array}{*{20}{c}}
      {,\,\,\,{\rm{if}} \,\,\, 0 < \tau _0^{\star} \le \frac{{{P_{avg}}}}{{{P_{peak}}}}}  \\
      {,\,\,\,\,\,\,\,\,\,\,\,\, {\rm{otherwise}} \,\,\,\,\,\,\,\,\,\, }  \\
      \end{array}} \right.,\,\,\,i = 1,\,\, \cdots ,\,\,K,}
   \end{equation}
   where $\tau_i^{\star}$, $i = 0, \,\, 1, \,\, \cdots, \,\, K$, are given in (\ref{Eq_Lemma_HD_tau_0}) and (\ref{Eq_Lemma_HD_tau_i}), and $z_i^* \ge 0$, $i = 1, \,\,  \cdots, \,\, K$, and $\lambda^* > 0$ are solutions of the following $(K+1)$ equations:
   \begin{equation}\label{Eq_Prop_HD_FunctionZ}
      {\ln \left( {1 + {z_i}} \right) - \frac{{{z_i}}}{{1 + {z_i}}} = \frac{\lambda^*}{{{\omega _i}}}\ln 2, \,\,\,\, i = 1, \,\, \cdots, \,\, K,}
   \end{equation}
   \begin{equation}\label{Eq_Prop_HD_SumCondition}
      {\sum\limits_{i = 1}^K {\frac{{{\alpha _i}}}{{{z_i}}} = \frac{1}{{{P_{avg}}}} - \frac{1}{{{P_{peak}}}}} .}
   \end{equation}
\begin{proof}
   Please refer to Appendix \ref{App_Proof_Prop_HD_OptSolution}.
\end{proof}

From Proposition 4.1, it is observed that the optimal transmit power of the H-AP in HD-WPCN is always $P^* = P_{peak}$, regardless of the optimal time allocations. In addition, we can compute ${\boldsymbol{\tau}}^{*}$ for (P5) efficiently as follows. First, we solve (P6) to obtain ${\boldsymbol{\tau}}^{\star}$ (one algorithm to solve (P6) is given in \cite{Ju}). If $\tau_0^{\star} \le \frac{P_{avg}}{P_{peak}}$, then the optimal time allocation solution for (P5) is $\boldsymbol{\tau}_i^* = \boldsymbol{\tau}_i^{\star}$, $i = 0, \,\, 1, \,\, \cdots, \,\, K$. Otherwise, if $\tau_0^{\star} > \frac{P_{avg}}{P_{peak}}$, then $\tau_0^* = \frac{P_{avg}}{P_{peak}}$, and $\tau_i^*$, $i = 1, \,\, \cdots, \,\, K$, can be obtained from (\ref{Eq_Prop_HD_Opt_UL_time}) via solving for $z_i^*$'s by a bisection search over $\lambda$ until both (\ref{Eq_Prop_HD_FunctionZ}) and (\ref{Eq_Prop_HD_SumCondition}) hold with equality.

Similar to the case of FD-WPCN with perfect SIC, we further investigate the optimal time and power allocation for HD-WPCN in Proposition 4.1. First, consider the special case of $P_{peak} = \infty$. In this case, it can be shown from (\ref{Eq_Prop_HD_Opt_DL_Time}) that $\tau_0^* = 0$ and thus $P^* \rightarrow \infty$ such that $\tau_0^* P^* \rightarrow P_{avg}$. In addition, it can be shown from (\ref{Eq_Lemma_HD_tau_i}) that $\tau_i^*$, $i = 1, \,\, \cdots, \,\, K$, increases with $\alpha_i$ or $\omega_i$ since $z_i^{\star}$ decreases with $\omega_i$ according to (\ref{Eq_Lemma_HD_FunctionZ}). Note that these observations have been similarly made for the FD-WPCN in the special case of $P_{peak} \to \infty$. Next, consider the general case of $P_{peak} < \infty$ and furthermore $0 < \tau_0^{\star} < \frac{P_{avg}}{P_{peak}}$. In this case, increasing $\tau_0$ above $\tau_0^{\star}$ reduces the WSR of HD-WPCN, although the amount of harvested energy of each user in DL WET and hence its transmit power in UL WIT increases with increasing $\tau_0$. This is because the resulting decrease of transmission time for UL WIT reduces the WSR more substantially than the improvement of WSR due to the increase of transmit power. At last, consider the case of $P_{peak} < \infty$ and $\tau_0^{\star} > \frac{P_{avg}}{P_{peak}}$. In this case, minimizing $\tau_0$, i.e., by setting $\tau_0^* = \frac{P_{avg}}{P_{peak}}$, maximizes the WSR since the time allocated for UL WIT is maximized for the given total energy sent by the H-AP, which is $P_{avg}$. For both the above two cases with $P_{peak} < \infty$, $\tau_i^*$, $i = 1, \,\, \cdots, \,\, K$, increases with $\alpha_i$ or $\omega_i$, similar to the case with $P_{peak} = \infty$.

\begin{figure}[!t]
   \centering
   \includegraphics[width=1.0\columnwidth]{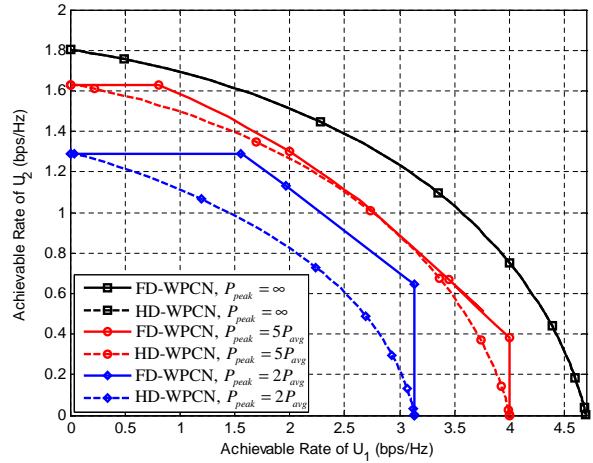}
   \caption{Rate region comparison between FD-WPCN with perfect SIC versus HD-WPCN.}
   \label{Fig_RateRegion_FD_vs_HD_No_SI}
\end{figure}

Fig. \ref{Fig_RateRegion_FD_vs_HD_No_SI} compares the achievable rate regions of FD-WPCN (assuming perfect SIC) versus HD-WPCN with $K = 2$ based on Propositions 3.1 and 4.1, respectively, with the same two-user channel setup as for Fig. \ref{Fig_RateRegion_No_SI} and three different peak power values, $P_{peak} = 2P_{avg}$, $5P_{avg}$, and $\infty$. It is worth noting that when $P_{peak} = \infty$, the achievable rate regions of both FD-WPCN and HD-WPCN are identical, since they have the same time and power allocation. However, when $P_{peak} < \infty$, the achievable rate region of HD-WPCN is observed to be smaller than that of FD-WPCN. This is because the H-AP in the FD-WPCN can broadcast all of its available energy to users, whereas that in the HD-WPCN in general cannot due to the total time constraint. Furthermore, $\tau_0^*$ in the FD-WPCN is in general smaller than that of HD-WPCN since the other $K$ slots in the FD-WPCN can also be used for DL WET whereas this is not possible for the HD-WPCN; as a result, more time can be allocated to users for UL WIT in the case of FD-WPCN, leading to a larger rate-region. Particularly, it is observed that when $P_{peak} = 2P_{avg}$, the achievable rate region of FD-WPCN is observed to be more notably larger than that of HD-WPCN. Therefore, the benefit achieved by employing FD H-AP over HD counterpart is more significant with smaller $P_{peak}$, but is less evident as $P_{peak}$ increases.

\section{Simulation Results}\label{Sec:SimulationResults}
In this section, we compare the achievable sum-rates of FD-WPCN with/without SI versus HD-WPCN under a practical system setup. The bandwidth is set as $1$MHz. The distance between the H-AP and user $U_i$, denoted by $D_i$, is assumed to be uniformly distributed within $5{\rm{m}} \le D_i \le 10 {\rm{m}}$, $i=1,\, \cdots, \, K$. For each user, the DL and UL channel power gains are modeled as $|h_{D,i}|^2 = 10^{-3}{\rho _{D,\,i}} { {D_i^{ - \alpha_D}}}$ and $|h_{U,i}|^2 = 10^{-3}{\rho _{U,\,i}} { {D_i^{ - \alpha_U}}}$, respectively, where the same pathloss exponents $\alpha_D = \alpha_U = 2$ are assumed. In addition, $\rho_{D,\,i}$ and $\rho_{U,\,i}$ represent the channel short-term fading in the DL and UL, respectively, which are both assumed to be Rayleigh distributed, i.e., $\rho_{D,\,i}$ and $\rho_{U,\,i}$ are independent exponential random variables with unit mean. Note that in the above channel model, a $30$dB average signal power attenuation is assumed at a reference distance of $1$m. The AWGN at the H-AP receiver is assumed to have a constant power spectral density of $-160$dBm/Hz. For all users, it is assumed that $\theta_i = 0.5$\footnote{Assuming that practical rectifier antenna is employed for wireless energy harvesting, the energy harvesting efficiency is typically in the range of $70$-$80\%$ \cite{Rectenna}. Given that ${\zeta _i} = 75\%$, $\theta_i = 0.5$ is obtained by assuming that $67\%$ of the harvested energy is used for WIT in the UL at $U_i$, i.e., $\eta_i = 2/3$, $i = 1, \,\, \cdots, \,\, K$.}, $\forall i = 1, \,\, \cdots, \,\, K$. We set $\Gamma = 9.8$dB assuming that an uncoded quadrature amplitude modulation (QAM) is employed with the required bit-error-rate (BER) of $10^{-7}$ \cite{Goldsmith}. Furthermore, to account for quantization error after ADC, we set $\beta = -60$dB as assumed in \cite{Day}. Finally, it is assumed that the mean-square error (MSE) of channel estimation error of the loopback channel is set to be $\varepsilon = -60$dB due to the large power of the loopback signal.

\begin{figure}[!t]
   \centering
   \includegraphics[width=1.0\columnwidth]{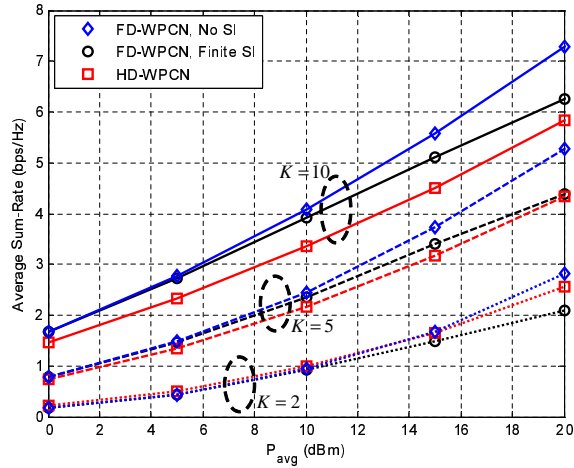}
   \caption{Average sum-rate vs. $P_{avg}$ with $K=10$, $P_{peak} = 2P_{avg}$, and $\varphi = - 60$dB.}
   \label{Fig_AvgRate_vs_Pavg_Various_K}
\end{figure}

Fig. \ref{Fig_AvgRate_vs_Pavg_Various_K} shows the average sum-rates of FD-WPCN versus HD-WPCN for different values of $P_{avg}$ in dBm by averaging over $1000$ randomly generated fading channel realizations, with $K=10$, $P_{peak} = 2P_{avg}$, and $\varphi = - 60$dB, i.e., $-60$dB SIC in analog domain\footnote{By current techniques, it has been reported that SI can be canceled up to $-81$dB in analog domain \cite{Sahai}.}. According to (\ref{Eq_Effective_SI}), the power of effective SI in this case is -123dB less than the power of the transmitted signal of the H-AP at a given time slot. As shown in Fig. \ref{Fig_AvgRate_vs_Pavg_Various_K}, the average sum-rate of FD-WPCN is always larger than that of HD-WPCN when SI is perfectly eliminated. In particular, FD-WPCN with perfect SIC outperforms HD-WPCN more considerably as the number of users in the network, $K$, increases. However, when SI in FD-WPCN is not perfectly cancelled, HD-WPCN outperforms FD-WPCN with small number of users in the network, e.g., $K = 2$. As $K$ increases, FD-WPCN even with finite SI can outperform HD-WPCN when $P_{avg}$ is sufficiently small or the FD-WPCN is less interference limited, e.g., the sum-rate of FD-WPCN with finite SI is larger than that of HD-WPCN when $0{\rm{dBm}} \le P_{avg} \le 25{\rm{dBm}}$ with $K =10$.

\begin{figure}[!t]
   \centering
   \includegraphics[width=1.0\columnwidth]{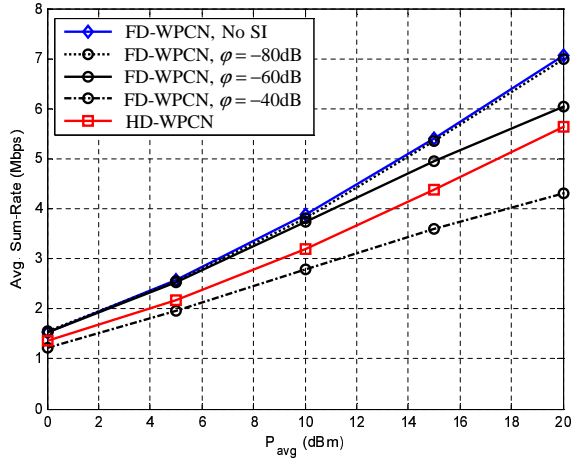}
   \caption{Average sum-rate vs. $P_{avg}$ for different values of $\varphi$ with $K=10$ and $P_{peak} = 2P_{avg}$.}
   \label{Fig_AvgRate_vs_Pavg_Various_SI}
\end{figure}

\begin{figure}[!t]
   \centering
   \includegraphics[width=1.0\columnwidth]{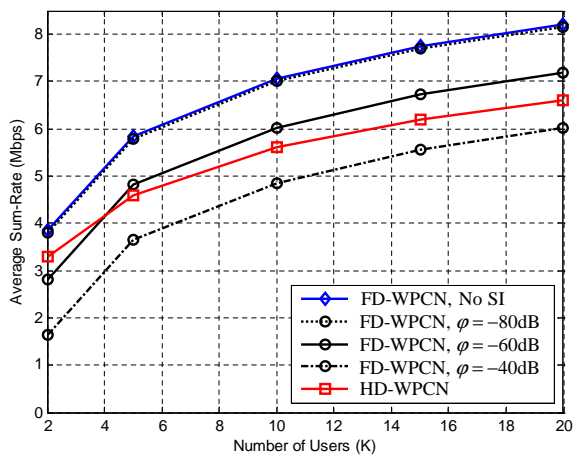}
   \caption{Average sum-rate vs. $K$ for different values of $\varphi$ with $P_{avg} = 20$dBm and $P_{peak} = 2P_{avg}$.}
   \label{Fig_AvgRate_vs_K_Various_SI}
\end{figure}

Next, by fixing $K = 10$, Fig. \ref{Fig_AvgRate_vs_Pavg_Various_SI} shows the average sum-rate comparison for different values of $\varphi$, which measures the effective loopback channel power after analog domain SIC. It is observed that when $\varphi = -80$dB, the sum-rate of FD-WPCN with imperfect SIC converges to that with perfect SIC, and also outperforms HD-WPCN. When $\varphi = -60$dB, the sum-rate of FD-WPCN with imperfect SIC is larger than that of HD-WPCN when $P_{avg} < 25$dBm, but becomes smaller when $P_{avg} \ge 25$dBm. Furthermore, the sum-rate of FD-WPCN with imperfect SIC is always smaller than that of HD-WPCN when $\varphi = -40$dB.

Fig. \ref{Fig_AvgRate_vs_K_Various_SI} shows the average sum-rates of FD- and HD-WPCNs over the number of users, $K$, for different values of $\varphi$. It is observed that the achievable sum-rates of both FD- and HD-WPCNs increase with $K$. In addition, when $\varphi = -80$dB, FD-WPCN with imperfect SIC is observed to have comparable achievable sum-rate to that with perfect SIC and also have larger sum-rate than HD-WPCN over all values of $K$. Furthermore, when $\varphi = -60$dB, it is observed that the average sum-rate of FD-WPCN with imperfect SIC is smaller than that of HD-WPCN when $K < 5$, but becomes larger when $K \ge 5$. However, the average sum-rate of FD-WPCN with imperfect SIC is observed to be always smaller than that of HD-WPCN when $\varphi = -40$dB.

\begin{figure}[!t]
   \centering
   \includegraphics[width=1.0\columnwidth]{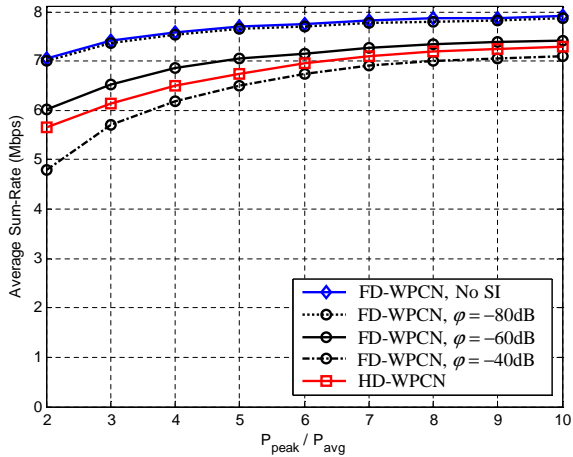}
   \caption{Average sum-rate vs. $P_{peak}/P_{avg}$ for different values of $\varphi$ with $K=10$ and $P_{avg} = 20$dBm.}
   \label{Fig_AvgRate_vs_Ppeak_Various_SI}
\end{figure}

At last, Fig. \ref{Fig_AvgRate_vs_Ppeak_Various_SI} shows the average sum-rate comparison for different values of $P_{peak}/P_{avg}$ with $K = 10$ and $P_{avg} = 20$dBm. It is observed that the achievable sum-rate gain of FD-WPCN without SI is more pronounced over HD-WPCN when $P_{peak}/P_{avg}$ is small, but the gap decreases with increasing $P_{peak}/P_{avg}$. In addition, when $P_{peak}/P_{avg} \to \infty$, it is also observed that the sum-rate of FD-WPCN becomes more comparable to that of HD-WPCN even with imperfect SIC.

\section{Conclusion} \label{Conclusion}
This paper studied optimal resource allocation in a new type of WPCN where full-duplex H-AP is employed, namely FD-WPCN. We proposed a new transmission protocol for the FD-WPCN which enables efficient simultaneous WET in DL and WIT in UL, over the same bandwidth. With the proposed protocol, we studied the joint time and power allocation in FD-WPCN to maximize the WSR in both the ideal case assuming perfect SIC and the practical case with finite residue SI. It is shown that to maximize the WSR of FD-WPCN, the optimal time and power allocation should optimally exploit the available multiuser channel diversity in the hybrid network. We also studied the optimal joint time and power allocation for a baseline HD-WPCN, and compared the achievable rates with FD-WPCN. Simulation results revealed that the FD H-AP is more beneficial than HD H-AP in WPCNs when the SI can be effectively cancelled, the number of users in the network is sufficiently large, and/or the peak transmit power constraint is more stringent as compared to the average transmit power at the H-AP.

\appendices

   \section{Proof of Lemma 3.1}\label{App_Proof_FD_WSR_Concavity}
   For any $i = 1, \,\, \cdots, \,\, K$, ${R_i^{(\rm{F-NoSI})}}\left( {\boldsymbol{\tau} ,{\bf{E}}} \right)$ given in (\ref{Eq_APP_FD_AchievableRate_No_SI_New}) is a perspective of $f_i \left( {\bf{E}} \right)$ \cite{Boyd}, where
   \begin{equation}\label{Eq_APP_Perspective}
      {f_i \left( {\bf{E}} \right) \buildrel \Delta \over =   \log_2 \left( {1 + \alpha_i \left( P_{avg} - E_i \right) } \right).}
   \end{equation}
   Note that $f_i \left( {\bf{E}} \right)$ in (\ref{Eq_APP_Perspective}) is a concave function of ${\bf{E}}$ because $f_i \left( {\bf{E}} \right)$ is a composition of a concave function ${{\hat f}_i}\left( x \right) = {\log _2}\left( {1 + {\alpha _i}x} \right)$ and an affine function $\tilde f_i \left( {\bf{E}} \right) = P_{avg} - E_i$, i.e. $f_i \left( {\bf{E}} \right) = {{\hat f}_i}\left( {\tilde f_i\left( {\bf{E}} \right)} \right)$. Since the perspective operation preserves concavity \cite{Boyd}, ${R_i^{(\rm{F-NoSI})}}\left( {\boldsymbol{\tau} ,{\bf{E}}} \right)$ is thus a jointly concave function of $\boldsymbol{\tau}$ and $\bf{E}$, $\forall i = 1, \,\, \cdots, \,\, k$. This thus completes the proof of Lemma 3.1.

   \section{Proof of Proposition 3.1}\label{App_Proof_Prop_FD_NoSI_OptSolution}
   Lagrangian given in (\ref{Eq_Lagrangian}) can be alternatively expressed as
   \begin{equation}\label{Eq_App_Lagrangian_i}
      {\mathcal L\left( {\boldsymbol{\tau} ,{\bf{E}},\lambda ,\mu } \right) =  \sum\limits_{i = 0}^K {{\mathcal L_i}\left( {{\tau _i},{E_i},\lambda ,\mu } \right)}  + \lambda  - \mu {P_{avg}}, }
   \end{equation}
   where ${{\mathcal L_i}\left( {{\tau _i},{E_i},\lambda ,\mu } \right)}$, $i = 0, \,\, 1, \,\, \cdots, \,\, K$ is given by
   \[
      {\mathcal L_i}\left( {{\tau _i},{E_i},\lambda ,\mu } \right) \,\,\,\,\,\,\,\,\,\,\,\,\,\,\,\,\,\,\,\,\,\,\,\,\,\,\,\,\,\,\,\,\,\,\,\,\,\,\,\,\,\,\,\,\,\,\,\,\,\,\,\,\,\,\,\,\,\,\,\,\,\,\,\,\,\,\,\,\,\,\,\,\,\,\,\,\,\,\,\,\,\,\,\,\,\,\,\,\,\,\,\,\,\,\,\,\,\,\,\,\,\,\,\,
   \]
   \begin{equation}\label{App_Lagrangian_i}
      {= \left\{ {\begin{array}{*{20}{c}}
      { - \lambda {\tau _0} + \mu {E_0}}  \\
      {{\omega _i}\hat R_i^{({\rm{F - NoSI}})}\left( {\boldsymbol{\tau} ,{\bf{E}}} \right) - \lambda \tau_i - \mu E_i}  \\
      \end{array}\begin{array}{*{20}{c}}
      {,\,\,\,\,\,\,\,\, i = 0\,\,\,\,\,\,\,}  \\
      {,\,\,{\rm{otherwise.}}}  \\
      \end{array}} \right.}
   \end{equation}
   Given $\lambda$ and $\mu$, $\mathcal G\left( {\lambda ,\mu } \right)$ in (\ref{Eq_DualFunction}) can be obtained by maximizing individual ${{\mathcal L_i}\left( {{\tau _i},{E_i},\lambda ,\mu } \right)}$, $\forall i = 0, \,\, 1, \,\, \cdots, \,\, K$, subject to $(\tau_i,E_i) \in \mathcal D$, since ${{\mathcal L_i}\left( {{\tau _i},{E_i},\lambda ,\mu } \right)}$ depends only on $\tau_i$ and $E_i$, i.e., by solving the following problem, $\forall i = 0, \,\, 1, \,\, \cdots, \,\, K$.
   \[
      \mathop {\max }\limits_{{\tau _i},\,\,{E_i}} \,\,\,{\mathcal L_i}\left( {{\tau _i},{E_i},\lambda ,\mu } \right) \,\,\,\,
   \]
   \[
      \,\,\,\,\,\, {\rm{s.t.}} \,\,\, 0 \le E_i \le P_{peak} \tau_i,
   \]
   \begin{equation}\label{Eq_App_Max_Lagrangian_i}
      0 \le \tau_i \le 1.
   \end{equation}

   We first consider the problem in (\ref{Eq_App_Max_Lagrangian_i}) for the case with $i = 1, \,\, \cdots, \,\, K$. Given $\lambda$, $\mu$, and $\tau_i$, we can obtain $E_i$ that maximizes ${{\mathcal L_i}\left( {{\tau _i},{E_i},\lambda ,\mu } \right)}$ in (\ref{App_Lagrangian_i}) by setting $\frac{\partial }{{\partial {E_i}}} = 0$, from which we have
   \begin{equation}\label{App_Differentiate_Ei}
      {\frac{{{\alpha _i}{\omega _i}}}{{1 + \frac{{{\alpha _i}}}{{{\tau _i}}}\left( {{P_{avg}} - {E_i}} \right)}} = \mu \ln 2.}
   \end{equation}
   Since $0 \le E_i \le P_{peak}\tau_i$ in $\mathcal D$ as shown in (\ref{Eq_FD_PeakPowerConstraint_New2}) and (\ref{Eq_FD_PositiveConstraint}), $E_i$ is given from (\ref{App_Differentiate_Ei}) as (\ref{Eq_Prop_FD_Opt_Ei}). It is shown from (\ref{App_Differentiate_Ei}) that $0 \le E_i \le P_{avg}$ if $\mu > 0$, $E_i \to -\infty$ if $\mu = 0$, and $E_i > P_{avg}$ if $\mu < 0$. In addition, given $\lambda$, $\mu$, and $E_i$, we can also find $\tau_i$ that maximizes ${{\mathcal L_i}\left( {{\tau _i},{E_i},\lambda ,\mu } \right)}$ in (\ref{App_Lagrangian_i}) by setting $\frac{\partial }{{\partial {\tau_i}}} = 0$, from which we have
   \begin{equation}\label{App_Differentiate_taui}
      {{\ln \left( {1 + {z_i}} \right) - \frac{{{z_i}}}{{1 + {z_i}}}} = \frac{\lambda \ln 2}{\omega_i},}
   \end{equation}
   where ${z_i} = \frac{{{\alpha _i}}}{{{\tau _i}}}\left( {{P_{avg}} - {E_i}} \right)$.
   Since $0 \le \tau_i \le 1$ in $\mathcal D$ as shown in (\ref{Eq_FD_PositiveConstraint}), $\tau_i$ is thus given as (\ref{Eq_Prop_FD_Opt_tau_i}) when $z_i$ is the solution of (\ref{App_Differentiate_taui}).

   Next, consider the case with $i = 0$. Since $0 \le E_0 \le P_{peak} \tau_0$, it is easily shown from (\ref{App_Lagrangian_i}) that $E_0$ maximizing ${{\mathcal L_0}\left( {{\tau _0},{E_0},\lambda ,\mu } \right)}$ is given as (\ref{Eq_Prop_FD_Opt_E0}). Note that when $\mu > 0$, $E_0$ is given by $E_0 = P_{peak}\tau_0$ and thus ${{\mathcal L_0}\left( {{\tau _0},{E_0},\lambda ,\mu } \right)} = \left( -\lambda + \mu P_{peak} \right)\tau_0$. Since $0 \le \tau_0 \le 1$, to maximize ${{\mathcal L_0}\left( {{\tau _0},{E_0},\lambda ,\mu } \right)}$ we have $\tau_0 = 1$ if $-\lambda + \mu P_{peak} > 0$ and $\tau_0 = 0$ otherwise. When $\mu \le 0$, $\tau_0 = 0$ to maximize ${{\mathcal L_0}\left( {{\tau _0},{E_0},\lambda ,\mu } \right)}$ since $E_0 = 0$ and thus ${{\mathcal L_0}\left( {{\tau _0},{E_0},\lambda ,\mu } \right)} = -\lambda \tau_0$. Therefore, $\tau_0$ maximizing ${{\mathcal L_0}\left( {{\tau _0},{E_0},\lambda ,\mu } \right)}$ is given by (\ref{Eq_Prop_FD_Opt_tau_0}).

   This completes the proof of Proposition 3.1.

   \section{Proof of Corollary 3.1}\label{App_Proof_Cor_FD_Max_WSR_Infinite_PPC}
   To prove Corollary 3.1, we consider the following two cases for any $i = 1 \,\, \cdots, \,\, K$: $E_i^* = 0$ and $E_i^* > 0$. First, when $E_i^* = 0$, it follows from (\ref{Eq_Prop_FD_Opt_tau_i}) that $\tau_i^* = \frac{\alpha_i}{z_i^*} P_{avg} > 0$. Next, when $E_i^* > 0$, from (\ref{Eq_Prop_FD_Opt_tau_i}) and (\ref{Eq_Prop_FD_Opt_Ei}) we have
   \begin{equation}\label{Eq_App_taui_1}
      \tau_i^* = {\frac{{{\alpha _i}}}{{z_i^*}}\left( {{P_{avg}} - E_i^*} \right)},
   \end{equation}
   \begin{equation}\label{Eq_App_Ei_1}
      {E_i^* = {P_{avg}} + \tau _i^*\left( {\frac{1}{{\alpha _i}} - \frac{{{\omega _i}}}{{{\mu ^*}\ln 2}}} \right),}
   \end{equation}
   where (\ref{Eq_App_taui_1}) is originated from the fact that $\tau_i^* < 1$. By substituting $\tau_i^*$ in (\ref{Eq_App_Ei_1}) by (\ref{Eq_App_taui_1}), it follows that
   \[
      E_i^*\left( {1 + \frac{{{\alpha _i}}}{{z_i^*}}\left( {\frac{1}{{{\alpha _i}}} - \frac{{{\omega _i}}}{{{\mu ^*}\ln 2}}} \right)} \right)
   \]
   \[
       = {P_{avg}}\left( {1 + \frac{{{\alpha _i}}}{{z_i^*}}\left( {\frac{1}{{{\alpha _i}}} - \frac{{{\omega _i}}}{{{\mu ^*}\ln 2}}} \right)} \right),
   \]
   from which we have $E_i^* = P_{avg}$. With $P_{peak} \to \infty$, therefore, we have $(\tau_i^*,E_i^*) = (0,P_{avg})$ or $(\frac{\alpha_i}{z_i^*}P_{avg},0)$.

   If there is an $i \in \left\{ {1,\cdots,K} \right\}$ such that $(\tau_i^*,E_i^*) = (0,P_{avg})$, it follows that $E_j^* = 0$, $\forall j=0, \,\, 1, \,\, \cdots, \,\, K$, $j \ne i$, to satisfy the sum-energy constraint in (\ref{Eq_FD_SumEnergyConstraint_new}). In addition, from (\ref{Eq_Prop_FD_Opt_tau_i}) we have $\tau_i^* = 0$ and $\tau_j = \frac{\alpha_i}{z_i^*}P_{avg}$, $\forall j = 1, \cdots, K$, $j \ne i$. In this case, FD-WPCN with $P_{peak} \to \infty$ becomes equivalent to $(K-1)$-user TDMA network consisting of $U_1, \cdots, U_{i-1}, U_{i+1}, \cdots, U_{K}$, where each user consumes a constant energy for information transmission and the channel of $U_j$ is given by $\alpha_j$. If $(\tau_i^*,E_i^*) = (\frac{\alpha_i}{z_i^*}P_{avg},0)$, $\forall i = 1, \cdots, K$, it then follows that $E_0^* = P_{avg}$ and $\tau_0^* = 0$. In this case, FD-WPCN with $P_{peak} \to \infty$ becomes equivalent to $K$-user TDMA network consisting of $U_i$'s, $\forall i=1,\cdots,K$, where each user consumes a constant energy for information transmission and the channel of $U_i$ is given by $\alpha_i$.

   It is well known that the achievable rate region of $(K-1)$-user TDMA network is a subset of $K$-user TDMA network. Therefore, $(\tau_0^*,E_0^*) = (0,P_{avg})$ and $(\tau_i^*,E_i^*) = (\frac{\alpha_i}{z_i^*}P_{avg},0)$, $\forall i = 1, \cdots, K$, is the optimal time and energy allocation solution for problem (P2) with $P_{peak} \to \infty$. Corollary 3.1 is thus proved.

   \section{Proof of Lemma 3.2}\label{App_Proof_FD_SI_Concavity}
   From (\ref{Eq_FD_AchievableRate_SI}), the Hessian of $R_i^{({\rm{F-SI}})} ( {\tau_i ,{P_i^{(k-1)}}} )$, $i = 1, \,\, \cdots, \,\, K$, is given by ${\nabla ^2}R_i^{({\rm{F-SI}})} ( {\tau_i ,{P_i^{(k-1)}}} ) = [ {d_{j,k}^{(i)}} ]$, $j = 0, \cdots, K$, $k = 0, \cdots, K$, where
   \begin{equation}\label{Eq_App_Concavity_SI}
      {d_{j,k}^{(i)} = \left\{ {\begin{array}{*{20}{c}}
      { - \frac{{C_i^2P_{avg}^2}}{{{\tau _i}{{\left( {{\tau _i} + {C_i}\left( {{P_{avg}} - P_i^{(k-1)}{\tau _i}} \right)} \right)}^2}}}}  \\
      0  \\
      \end{array}} \right.\begin{array}{*{20}{c}}
      {,\,\,\,j = k = i}  \\
      {,\,\,{\rm{otherwise}},}  \\
      \end{array}}
   \end{equation}
   with $C_i = \frac{\theta_i H_i}{\Gamma \left({{\gamma P_i^{(k-1)}} + {{\sigma }^2}}\right)}$. We thus have ${{\bf{v}}^T}{\nabla ^2}R_i^{({\rm{F-SI}})} ( {\tau_i ,{{P}^{(k-1)}}} ){\bf{v}} \le 0$ for any given ${\bf{v}} = [v_0, \,\, \cdots, \,\, v_K]^T$. This thus proves Lemma 3.2.

   \section{Proof of Proposition 3.2}\label{App_Proof_Prop_FD_SI_OptTime}
   Given $\lambda$, $\mu$, and ${\bf{P}}^{(k-1)}$, $\mathcal L^{(\rm{F-SI})} (\boldsymbol{\tau}, \lambda, \mu)$ can be expressed as
   \begin{equation}\label{Eq_App_Finite_SI_Lagrangian}
      \mathcal L^{(\rm{F-SI})} (\boldsymbol{\tau}, \lambda, \mu) = \sum\limits_{i = 0}^K \mathcal L_i^{(\rm{F-SI})} (\tau_i, \lambda, \mu),
   \end{equation}
   where $L_i^{(\rm{F-SI})} (\tau_i, \lambda, \mu)$ is given by
   \begin{equation}\label{Eq_App_Finite_SI_Lagrangian_i}
      L_i^{(\rm{F-SI})} (\tau_i, \lambda, \mu) \,\,\,\,\,\,\,\,\,\,\,\,\,\,\,\,\,\,\,\,\,\,\,\,\,\,\,\,\,\,\,\,\,\,\,\,\,\,\,\,\,\,\,\,\,\,\,\,\,\,\,\,\,\,\,\,\,\,\,\,\,\,\,\,\,\,\,\,\,\,\,\,\,\,\,\,\,\,\,\,\,\,\,\,\,\,\,\,\,\,\,\,\,\,\,\,\,\,\,\,\,\,\,\,\,\,\,\,\,\,\,\,
   \end{equation}
   \[
      =\left\{ {\begin{array}{*{20}{c}}
      { - \left( {\lambda  + P_i^{(k - 1)}\mu } \right){\tau _0}}  \\
      {{\omega _i}{\tau _i}{{\log }_2}\left( {1 + {C_i}\frac{{{P_{avg}} - P_i^{(k - 1)}{\tau _i}}}{{{\tau _i}}}} \right) }  \\
      {- \left( {\lambda  + P_i^{(k - 1)}\mu } \right){\tau _i} \,\,\,\,\,\,\,\,\,\,\,\,\,\,\,\,\,\,\,\,\,\,\,\, } \\
      \end{array}\,\,\begin{array}{*{20}{c}}
      {, \,\,\,\,\,\,\,\,\,\,\, i = 0 \,\,\,\,\,\,\,\,\,\,\,\,\,}  \\ {} \\
      {, \,\,\, i = 1, \cdots ,K.}  \\ {} \\
      \end{array}} \right.
   \]
   Similar to the proof of Proposition 3.1, $\mathcal L^{(\rm{F-SI})} (\boldsymbol{\tau}, \lambda, \mu)$ is maximized by maximizing individual $L_i^{(\rm{F-SI})} (\tau_i, \lambda, \mu)$, $i = 1, \,\, \cdots, \,\, K$. We first consider the case with $i = 0$. Since $0 \le \tau_0 \le 1$, from (\ref{Eq_App_Finite_SI_Lagrangian_i}) we have
   \begin{equation}\label{Eq_App_Finite_SI_tau0}
      {{{\bar \tau }_0} = \left\{ {\begin{array}{*{20}{c}}
      0  \\
      1  \\
      \end{array}\,\,\begin{array}{*{20}{c}}
      {,\,\,\,\lambda  + P_i^{(k - 1)}\mu  \ge 0}  \\
      {,\,\,\,\,\,\,\,\,\,\,\, {\rm{otherwise}}{\rm{.}} \,\,\,\,\,\,\,\,}  \\
      \end{array}} \right.}
   \end{equation}

   Next, consider the case with $i = 1, \,\, \cdots, \,\, K$. We can find $\bar{\tau}_i$ that maximizes $L_i^{(\rm{F-SI})} (\tau_i, \lambda, \mu)$ by setting ${\left. {\frac{\partial }{{\partial {\tau _i}}}L_i^{(\rm{F-SI})} (\tau_i, \lambda, \mu)} \right|_{{\tau _i} = {{\bar \tau }_i}}} = 0$, from which we have
   \begin{equation}\label{Eq_App_Finite_SI_KKT}
      {\bar f\left( z_i \right) = \ln \left( {1 + {z_i}} \right) - \frac{z_i + {{C_i}P_i^{(k - 1)}}}{{1 + {z_i}}} = \frac{{\left( {\lambda  + P_i^{(k - 1)}\mu } \right)\ln 2}}{{{\omega _i}}},}
   \end{equation}
   where ${z_i} = {C_i}\frac{{{P_{avg}} - P_i^{(k - 1)}{\tau _i}}}{{{\tau _i}}}$. When $z_i^{\star}$ is the solution of (\ref{Eq_App_Finite_SI_KKT}), we have (\ref{Eq_Prop_FD_SI_OptTime}) in Proposition 3.2.

   Note that $\bar f (z_i)$ given in (\ref{Eq_App_Finite_SI_KKT}) (also in (\ref{Eq_Function_bar_z})) is a monotonically increasing function of $z_i$, which has the minimum value $-C_i P_i^{(k-1)}$ at $z_i = 0$. Therefore, there is no $z_i^{\star}$ which is solution of (\ref{Eq_App_Finite_SI_KKT}) if $-C_i P_i^{(k-1)} > \frac{{\left( {\lambda  + P_i^{(k - 1)}\mu } \right)\ln 2}}{{{\omega _i}}}$. In addition, if $-C_i P_i^{(k-1)} \le \frac{{\left( {\lambda  + P_i^{(k - 1)}\mu } \right)\ln 2}}{{{\omega _i}}} < 0$, we can find $z_i^{\star}$ being a solution of (\ref{Eq_App_Finite_SI_KKT}) and thus non-zero $\bar \tau_i$, but $\bar \tau_0 = 1$ as shown in (\ref{Eq_App_Finite_SI_tau0}). This always violate the sum-time constraint in (\ref{Eq_FD_SumTime_NewNew}). Therefore, it should be satisfied that $\frac{{\left( {\lambda  + P_i^{(k - 1)}\mu } \right)\ln 2}}{{{\omega _i}}} \ge 0$, i.e., $( {\lambda  + P_i^{(k - 1)}\mu } )\ln 2 \ge 0$, with $\bar \tau_0 = 0$.

   This thus completes the proof of Proposition 3.2.

   \section{Proof of Proposition 4.1}\label{App_Proof_Prop_HD_OptSolution}
   We first consider the case where $0 < \tau_0^{\star} < \frac{P_{avg}}{P_{peak}}$. In this case, it is evident that the maximum WSR is obtained by setting $P^* = P_{peak}$ and ${\boldsymbol{\tau}}^* = {\boldsymbol{\tau}}^{\star}$ without violating the constraints in (\ref{Eq_HD_AvgPower}) and (\ref{Eq_HD_PeakPower}), since ${R_i^{(\rm{H})}}\left( {\boldsymbol{\tau} , P} \right)$ is a monotonically increasing function of $P$ as shown in (\ref{Eq_HD_AchievableRate}) and further increase of $\tau_0$ will decrease the WSR.

   Next, consider the other case where $\tau_0^{\star} > \frac{P_{avg}}{P_{peak}}$. In this case, $\tau_0^{\star}$ cannot be the optimal time allocation to DL WET since the average power constraint in (\ref{Eq_HD_AvgPower}) is violated by setting $P = P_{peak}$. It is worth noting that for any given $\frac{P_{avg}}{P_{peak}} \le \tau_0 \le 1$, the amount of harvested energy at a user does not change when $P$ is set to be $P = P_{avg} / \tau_0$ to satisfy the average power constraint in (\ref{Eq_HD_AvgPower}). For a given amount of harvested energy, the WSR decreases with increasing $\tau_0$ since ${R_i^{(\rm{H})}}\left( {\boldsymbol{\tau} , P} \right)$ is monotonically increasing function of $\tau_i$ and at least one $\tau_i$ should be decreased to satisfy the sum-time constraint in (\ref{Eq_HD_SumTime}). Therefore, the maximum WSR is obtained by setting $P^* = P_{peak}$ and $\tau_0^* = \frac{P_{avg}}{P_{peak}}$. In this case, problem (P3) can be modified as
   \[
      {\mathop {\max }\limits_{\tau_1, \,\, \cdots, \,\, \tau_K} \, \sum\limits_{i = 1}^K {{\omega _i} {\tau _i}{\log _2}\left( {1 + {\alpha _i}\frac{{{P_{avg}}}}{{{\tau _i}}}} \right) }}
   \]
   \begin{equation}\label{Eq_App_Modified_SumTimeConstraint}
      {{\rm{s.t.}}\,\,\,\,\,\, \sum\limits_{i = 1}^K {{\tau _i}}  \le 1 - \frac{P_{avg}}{P_{peak}}, \,\,\,\,\,\,\,\,\,\,\, }
   \end{equation}
   \[
      { \,\,\,\,\,\,\,\,\,\,\,\,\, \tau_i  \ge 0, \,\,\,\, i = 1, \,\, \cdots \,\, K, }
   \]
   which can be shown to be convex with strong duality. Therefore, the following KKT conditions should be satisfied by the optimal solution:
   \begin{equation}\label{Eq_App_HD_FunctionZ}
      {\ln \left( {1 + {\frac{\alpha_i}{\tau_i^*} P_{avg}}} \right) - \frac{{{\frac{\alpha_i}{\tau_i^*} P_{avg}}}}{{1 + {\frac{\alpha_i}{\tau_i^*} P_{avg}}}} = \frac{\lambda^*}{{{\omega _i}}}\ln 2, }
   \end{equation}
   \begin{equation}\label{Eq_App_HD_SumCondition}
      {\sum\limits_{i = 1}^K {\frac{{{\alpha _i}}}{{{\frac{\alpha_i}{\tau_i^*} P_{avg}}}} = \frac{1}{{{P_{avg}}}} - \frac{1}{{{P_{peak}}}}} ,}
   \end{equation}
   where $\lambda^*$ denotes the optimal Lagrange multiplier associated with the constraint in (\ref{Eq_App_Modified_SumTimeConstraint}). We obtain (\ref{Eq_Prop_HD_FunctionZ}) and (\ref{Eq_Prop_HD_SumCondition}) by changing variables as $z_i^* = \frac{\alpha_i}{\tau_i^*} P_{avg}$, $i = 1, \,\, \cdots, \,\, K$ and $\tau_i^*$ in (\ref{Eq_Prop_HD_Opt_UL_time}) is then obtained from (\ref{Eq_App_HD_SumCondition}).

   Combining both the parts above, the proof of Proposition 4.1 is thus completed.

\end{document}